\documentclass[12pt]{article}
\usepackage{amssymb,amsmath}
\usepackage{graphicx,epsfig}
\usepackage{caption}
\usepackage{subcaption}
\usepackage{enumerate}
\usepackage{epstopdf}
\usepackage{xcolor}
\usepackage{float}
\usepackage{rotating}
\usepackage{tikz}
\usepackage{dsfont}
\usepackage{natbib}
\usepackage{comment}
\usepackage{tabularx}
\usepackage{hyperref}
\usepackage{upgreek}
\usepackage[margin=3cm]{geometry}
\usepackage[english]{babel}
\usepackage[linesnumbered,lined,boxed,commentsnumbered]{algorithm2e}
\RestyleAlgo{ruled}
\SetKwComment{Comment}{/* }{ */}
\hypersetup{
    colorlinks=true,
    linkcolor=cyan,
    filecolor=cyan,      
    urlcolor=cyan,
    citecolor = cyan
    }
\graphicspath{{Rips sequences/}}

 \numberwithin{equation}{section}

\begin{document}
\newcounter{eqnarray}
\newtheorem{t1}{Theorem}[section]
\newtheorem{d1}{Definition}[section]
\newtheorem{c1}{Corollary}[section]
\newtheorem{l1}{Lemma}[section] \newtheorem{r1}{Remark}[section]
\newtheorem{e1}{Example}[section]
\newtheorem{p1}{Proposition}[section]

\title{Testing Homological Equivalence Using Betti Numbers : 
Probabilistic Properties}
\author{\small 
	Satish Kumar \\
	\small IIT Kanpur\\
	\small Department of Mathematics and Statistics \\
	\small  Kanpur 208016, India\\
	{\small email: satsh@iitk.ac.in }\\
	\and
	\small Subhra Sankar Dhar \\
	\small  IIT Kanpur\\
	\small   Department of Mathematics and Statistics \\
	\small Kanpur 208106, India\\
	{\small email: subhra@iitk.ac.in}\\
}

\maketitle

\textbf{Abstract:} In this article, we propose a novel one-sample test to check whether the support of the unknown distribution generating the data is homologically equivalent to the support of some specified distribution or not OR using the corresponding two-sample test, one can test whether the supports of two unknown distributions are homologically equivalent or not. In the course of this study, test statistics based on the Betti numbers are formulated, and the consistency of the tests is established under the critical and the supercritical regimes. Since these results are concerned with algebraic topology, we have reviewed the concepts from algebraic topology, and asymptotic results for the Betti numbers. Moreover, some simulation studies are conducted and results are compared with the existing methodologies such as Robinson's permutation test and the test based on mean persistent landscape functions. Furthermore, the practicability of the tests is shown on two well-known real data sets.    

{\bf keywords:} Topological data analysis, Simplicial complex, Homology, Betti numbers, Persistent homology, Persistent Betti numbers.


\section{Introduction}
Topological data analysis (TDA) is an emerging field in Statistics and data analysis as it can reveal many stimulating insights from complex data sets when the conventional techniques of data analysis are inadequate in analyzing the features of those data. In view of this, TDA is widely used in a diverse range of domains including biological sciences (\cite{Covid19(2022)}), finance (\cite{Finance2018}), astronomy (\cite{Astronomy(2016)}) and many more. Though many issues related to TDA have not yet been explored in the literature, for the recent development of the literature in TDA, see the articles by  \cite{doi:10.1146/annurev-statistics-031017-100045}, \cite{RobustCh(2017)}, \cite{FasyCI(2014)}, \cite{chazal:GeometricInference-2009}, \cite{GoF:Biscio(2014)}, \cite{Scandinavian} \cite{Bubenik:Statistical-2007}, \cite{Niyogi(2011)}, \cite{Landscape(2015)}, \cite{PDconvergence(2014)} and \cite{StochasticL(2014)}. In this article, we attempt to work on certain statistical testing of hypothesis problems using homology-a toolkit in TDA.

\subsection {Homology and Related Concepts} Suppose that we observe a random sample $\mathds{X} = \{X_{1}, X_{2},\ldots X_{n}\}$, where $X_i$ ($i = 1, \ldots, n$) is associated with some probability measure $\mathds{P}$ supported on a compact set $\mathcal{X}\subset\mathds{R}^d$, d $\geqslant$ 1, and the goal is to infer the topology of $\mathcal{X}$ from the observed data $\mathds{X}$. Note that one cannot consider $\mathds{X}$ as a topological space to estimate the topology of $\mathcal{X}$, since the topology of $\mathds{X}$ is trivial, and hence, $\mathds{X}$ does not contain any useful topological information. Therefore, one needs to find a way to transform $\mathds{X}$ into richer topological spaces that contain useful information about $\mathcal{X}$. The notion of a simplicial complex, from algebraic topology, plays a fundamental role in this regard. Simplicial complexes provide a way to convert the discrete set of points into richer geometric objects that contain useful information about the space underlying the data. Thus, the first step towards calculating the topology of $\mathcal{X}$ from the data is to convert the set $\mathds{X}$ into a simplicial complex, and there are various ways of constructing simplicial complexes from the data. In this article, we are mainly concerned with the $\check{C}$ech complex and the Vietoris-Rips complex that allows us to define and compute the homology efficiently.
Precisely speaking, this article concerns the topology of a topological space that can be approximated by the simplicial complex, and we are interested in topological properties in terms of homology. The homology of a simplicial complex is also referred to as simplicial homology.

Strictly speaking, Homology is a concept from algebraic topology that distinguishes two topological spaces based on the number of connected components and holes in a space.
Homology characterizes sets based on the connected components and holes in higher dimensions. In particular, $0^{th}$ order homology corresponds to the number of connected components of the set, which is the same as the number of clusters in the statistical sense.
Homology admits group structure and the rank of homology groups, are important topological invariants in TDA, also referred to as Betti numbers. This article is concerned with testing one sample and two sample homological equivalence using the calculated Betti numbers from the data since the Betti numbers characterize the homological equivalence.

\subsection{Hypothesis and Statistical Importance} From the statistical point of view, we develop one sample and two sample homological equivalence, which are as follows. For the one-sample problem, we investigate whether the support of the unknown distribution is homologically equivalent to the support of the specified distribution or not, and in the two-sample problem, we want to know whether the support of two unknown distributions is homologically equivalent or not. The statistical importance of such testing of hypothesis problems is many-fold. For instance, generating data from a uniform distribution with rectangular support can be carried out by a certain homeomorphic transformation of the data obtained from a uniform distribution with circular support. Besides, homological equivalence gives us insight into the clustering structure of the data, and hence, in order to know whether two data sets are forming two different clusters or not, the aforementioned two sample problems can be a useful device. Moreover, it is needless to mention that homological non-equivalence of the supports of the two distributions implies that those two distributions are not the same, and this fact can be an effective toolkit for goodness of fit problems and many other statistical methodologies.  In the course of this study, we calculate Betti numbers associated with a $\check{C}$ech or Vietoris rips complex constructed from the observed data. We propose test statistics based on the Betti numbers, and the consistency of the tests is also established. As mentioned earlier, this test is expected to be effective in practice as homology of the support of the distribution often has a substantial impact in exploratory data analysis, cluster analysis, reduction of dimension, and many other statistical applications. In other words, homological information enables us to extract surprising insights from complex data sets, which can further be used in data analysis for knowledge discovery. 

\subsection{Outline}
The rest of the article is organized as follows. As the concepts related to homology may be new to the readers, Section \ref{Preliminaries} provides the necessary background on algebraic topology needed to comprehend this article such as homology, Betti number, persistent homology, etc. Section \ref{Problem Formulation} formulates the research problem along with the consistency of the test. In Section \ref{Simulated data study}, we have conducted a simulated data study to examine the theoretical results. Section \ref{Conclusion} consists of a few concluding remarks. Finally, the Appendix (i.e., Section \ref{Appendix}) contains technical details.

{\it The R-codes of all numerical studies are available at    \url{https://github.com/satsh636/TDA.git}.}

\section{Preliminaries}\label{Preliminaries}

Given a random sample $X_{1},\ldots,  X_{n}$ from an unknown probability measure $\mathds{P}$, one may want to extract geometric features of the data to understand the inherent structure of the data. In TDA, one can extract topological features of the data. Strictly speaking, one may want to know the topology of the continuous object that underlies the data, i.e., topology of the support of the distribution. We now give the formal definition of topology and related notions.

\begin{d1}
\textbf{Topology:} Given a set of points, $\mathds{X}$, topology of the set $\mathds{X}$ denoted as $\tau$, is a collection of subsets of $\mathds{X}$ which satisfies the following properties:

\begin{itemize}
     
    \item Both the empty set $\phi$ and $\mathds{X}$ are in $\tau$.
    
    \item $\tau$ is closed under union.
    
    \item $\tau$ is closed under finite intersection.

\end{itemize}

\begin{r1}
If $\tau = \{\phi,\mathds{X}\}$, then $\tau$ is called \textbf{trivial topology.}
\end{r1}
\end{d1}

\begin{d1}
 
 \textbf{Topological space:} A topological space is a set $\mathds{X}$ equipped with its topology $\tau$, i.e., ($\mathds{X}, \tau$) is a topological space.
 
 \begin{r1}
 For any arbitrary set $\mathds{X}$, the topological space is generally denoted by the set $\mathds{X}$ itself.
 \end{r1}
 
 \begin{e1}
 
 Topology of the set of real numbers $\mathds{R}$ is the collection of all open sets $\mathcal{O}$ and the topological space is ($\mathds{R}, \mathcal{O}$). 
 
 \end{e1}
 
 \end{d1}

\begin{d1}
\textbf{Hausdorff space:} A topological space $\mathds{X}$ is said to be Hausdorff space if for any two distinct points $x$ and $y$ in $\mathds{X}$, there exists a neighbourhood $N_{x}$ of $x$ and a neighbourhood $N_{y}$ of $y$ such that $N_{x}\cap N_{y} = \phi$.
\end{d1}

\begin{d1}

\textbf{Homeomorphism:} 
Two topological spaces, $\mathds{X}$ and $\mathds{Y}$, are said to be topologically same or \textbf{homeomorphic} to each other if there exist a continuous bijective map with continuous inverse, such a map is called homeomorphism between the topological spaces $\mathds{X}$ and $\mathds{Y}$.

\end{d1}

\begin{d1}
\textbf{Manifold:} A manifold of dimension d or a d-manifold is a Hausdorff space in which each point has an open neighborhood
homeomorphic to the Euclidean space $\mathds{R}^d$.   
\end{d1}

One can distinguish two topological spaces by the property of homeomorphism. A property of a topological space that is invariant under homeomorphism, is called a topological invariant. If two topological spaces are homeomorphic to each other, then topological invariants associated with them coincide but converse is not true.
In general, classifying a topological space up to homeomorphism is not always possible. In TDA, one usually resorts to the homology of a topological space in order to distinguish two topological spaces. In particular, we are interested in simplicial homology which is defined for spaces that can be approximated through points, lines, triangles and its higher dimensional generalizations, and these are called simplicial complexes. In the following two subsections, we shall briefly define the concept of simplicial complex and simplicial homology. For details, one may refer to \cite{Munkres(1984)}. 

\subsection{Simplicial complex} \label{Simplicial complex}

In this section, we shall define the notion of a simplicial complex which belongs to an important class of topological spaces that we are interested in this article.

\begin{d1}
 
  \textbf{Complex:} A complex is a space that is constructed from a union of simple pieces (see below), if the pieces are topologically easily tractable and their common intersections are lower dimensional pieces of the same kind. ( \cite{Edelsbrunner(2010)})

  \noindent In particular, if a complex is made up of simple pieces like, points, line, edges, triangles and their higher dimensional analogues, then it is called a \textbf{Simplicial Complex.} Elements of a simplicial complex are called simplices. Points are 0-simplices, also called \textbf{vertices}; lines are 1-simplices, also called \textbf{edges}; triangles are 2-simplices and so on. 

\end{d1}

\begin{d1}

\textbf{Affine independence:} A collection of points $\{x_{0},x_{1},\ldots x_{n}\}$ is said to be affinely independent if the points $\{x_{1}-x_{0}, x_{2}-x_{0}\ldots x_{n}-x_{0}\}$ are linearly independent.
\end{d1}

\begin{d1}
\textbf{Simplex:} Suppose that $A = \{x_{0},x_{1},\ldots x_{n}\}$ is a set of affinely independent points in $\mathds{R}^d, d \geq 1$. An \textbf{n-simplex} or \textbf{n-dimensional simplex} spanned by $\{x_{0},x_{1},\ldots x_{n}\}$, denoted by $\sigma$ and defined as the set of convex combinations of $x_{0},x_{1},\ldots x_{n}$, i.e.,$$ \sigma = \left\{z\in \mathds{R}^d:z=\sum\limits_{i=0}^n \alpha_{i} x_{i} \text{ such that } \sum\limits_{i=0}^n \alpha_{i}=1;\alpha_{i} \geq 0 \text{ for all i = 0,1, }\ldots n\right\} $$

\noindent Here, the points $x_{0},x_{1},\ldots x_{n}$ are called vertices of $\sigma$. Simplices spanned by the subsets of A are called \textbf{faces} of $\sigma$.
\end{d1}

\begin{d1}
\textbf{ Simplicial complex:} A simplicial complex is a finite collection of simplices, denoted by $\mathcal{K}$, in $\mathds{R}^d, d \geq 1$ which satisfies the following condition:

\begin{itemize}
    
    \item If $\tau$ is any face of a simplex $\sigma$ in $\mathcal{K}$, then $\tau \in \mathcal{K}$.
    
    \item  If $\tau$ and $\sigma$ are two simplices in $\mathcal{K}$, then $\tau\bigcap\sigma = \phi$ or $\tau\bigcap\sigma$ is common face of both $\tau$ and $\sigma$.

\end{itemize}

  \noindent The simplicial complexes defined in this way are also referred as \textbf{geometric simplicial complexes}. Note that $\mathcal{K}$ can also be regarded as a topological space through its \textbf{underlying space} which is the union of its simplices. The union of simplices of $\mathcal{K}$ is a subset of $\mathds{R}^{d}$ that inherits topology from $\mathds{R}^{d}$. The simplices of $\mathcal{K}$ are called the \textbf{faces} of $\mathcal{K}$. The \textbf{dimension} of $\mathcal{K}$ is the largest dimension of its simplices. 

\end{d1}

\begin{d1}

\textbf{Subcomplex:} A subset of $\mathcal{K}$, which is itself a simplicial complex, is called a subcomplex of $\mathcal{K}$. In particular, a subcomplex, which contains all the simplices of $\mathcal{K}$ of dimension at most k, is called \textbf{k-skeleton} of $\mathcal{K}$. 
\end{d1}

In particular, 1-skeleton of a simplicial complex is same as \textbf{geometric graph}. Thus, simplicial complexes are higher dimensional generalisations of geometric graphs. Therefore, connectivity properties of a simplicial complex is same as its 1-skeleton, which is a geometric graph. Now, observe that construction of a simplicial complex involves specifying all the simplices and ensuring that simplices should intersect in a specified manner which is difficult in practice since it involves complicated geometric details. Therefore, in practice, one may work with the simplicial complexes which are fully characterized by only the list of its simplices, this leads to the notion of abstract simplicial complex.

\begin{d1}
\textbf{Abstract simplicial complex:}  Given a finite 
set $A = \{x_{0}, x_{1}, \ldots x_{n}\}$, a non-empty and finite collection of subsets of A, denoted by  $\mathcal{A}$ , is called an abstract simplicial complex if the following conditions are satisfied:

\begin{itemize}
    
    \item The elements of A belong to $\mathcal{A}$.
    
    \item If $\tau\in\mathcal{A}$ and $\sigma\subseteq\tau$, then $\sigma\in\mathcal{A}$.
    
\end{itemize}
\end{d1}

Note that given an abstract simplicial complex $\mathcal{A}$, one can always construct a geometric simplicial complex $\mathcal{K}$. Moreover, one can associate a topological space with the abstract simplicial complex $\mathcal{A}$ (denoted as $|\mathcal{A}|$) such that $|\mathcal{A}|$ is homeomorphic to the underlying space of $\mathcal{K}$, such a $\mathcal{K}$ is referred as \textbf{geometric realization} of $\mathcal{A}$. It can also be verified that any n-dimensional abstract simplicial complex can be realized as a geometric simplicial complex in $\mathds{R}^{2n+1}$. Therefore, one can consider abstract simplicial complexes as a topological space from which one can derive the topological properties.  

\begin{e1}
Given a set $A=\{a,b,c\}$, $\mathcal{A}= \{\{a\},\{b\},\{c\},\{a,b\},\{b,c\},\{c,a\}\} $ is an abstract simplicial complex representing the boundary of the triangle shown below: 
\begin{center}
\begin{tikzpicture}
        \draw (0,0)node[anchor=north]{$a$}--(4,0)node[anchor=north]{$b$}--(2,4)node[anchor=south]{$c$}--cycle; 
        
    \end{tikzpicture}
    \end{center}
    
    \noindent If the 2-simplex $\{a,b,c\}$ is added to the abstract simplicial complex $\mathcal{A}$, i.e.,  
    $$\mathcal{A} = \{\{a\},\{b\},\{c\},\{a,b\},\{b,c\},\{c,a\},\{a,b,c\}\},$$ 
    then it represent the triangle shown below:
     
     \begin{center}
\begin{tikzpicture}
        \draw (0,0)node[anchor=north]{$a$}--(4,0)node[anchor=north]{$b$}--(2,4)node[anchor=south]{$c$}--cycle; 
        \fill[black!30!white] (0,0)node[anchor=north]{$a$}--(4,0)node[anchor=north]{$b$}--(2,4)node[anchor=south]{$c$}--cycle;
        \draw (0,0)node[anchor=north]{$a$}--(4,0)node[anchor=north]{$b$}--(2,4)node[anchor=south]{$c$}--cycle;
    \end{tikzpicture}
    \end{center}
    
\end{e1}

\noindent \textbf{Data to abstract simplicial complex:}

\noindent Suppose that we have observed a random sample, $\mathds{X} = \{x_{1}, x_{2}, \ldots,  x_{n}\}$, of size n, from the unknown probability measure $\mathds{P}$ supported on the unknown manifold $\mathcal{X}\subseteq \mathds{R}^d, d\geq 1$, and one may want to estimate the topology of $\mathcal{X}$ from the observed sample $\mathds{X}$. Note that the set $\mathds{X}$ itself does not provide any useful topological information of $\mathcal{X}$, since the topology of $\mathds{X}$ is trivial. Since topology is a characteristics of a continuous space, therefore, in order to extract substantial topological information from $\mathds{X}$, one needs to connect the data points that are close or similar to each other in some sense so that one can have some idea about the continuous space underlying the data. 

\begin{e1}
Suppose that we are given 20 sample points, from a unit circle, drawn according to uniform probability measure. Then, the goal is to connect these data points so that one can estimate the topological properties of the underlying continuous space, that is the circle in this example (see Figure \ref{fig:circle})
\begin{figure}[!ht]
		\centering
		\begin{subfigure}[b]{0.49\linewidth}
			\includegraphics[width=\linewidth, height = 3in]{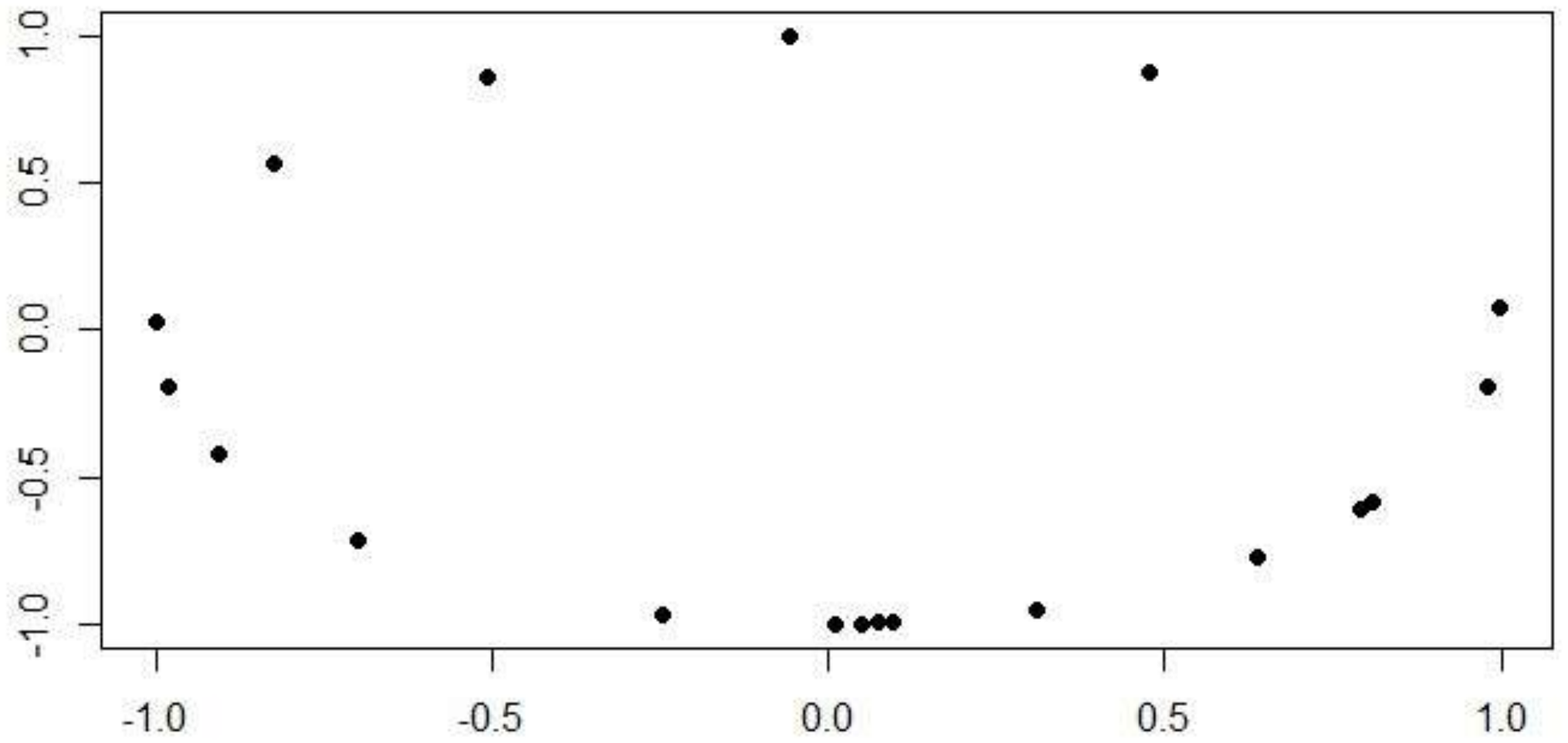}
			\caption{observed data }
		\end{subfigure}
	\hfill
		\begin{subfigure}[b]{0.49\linewidth}
			\includegraphics[width=\linewidth, height = 3in]{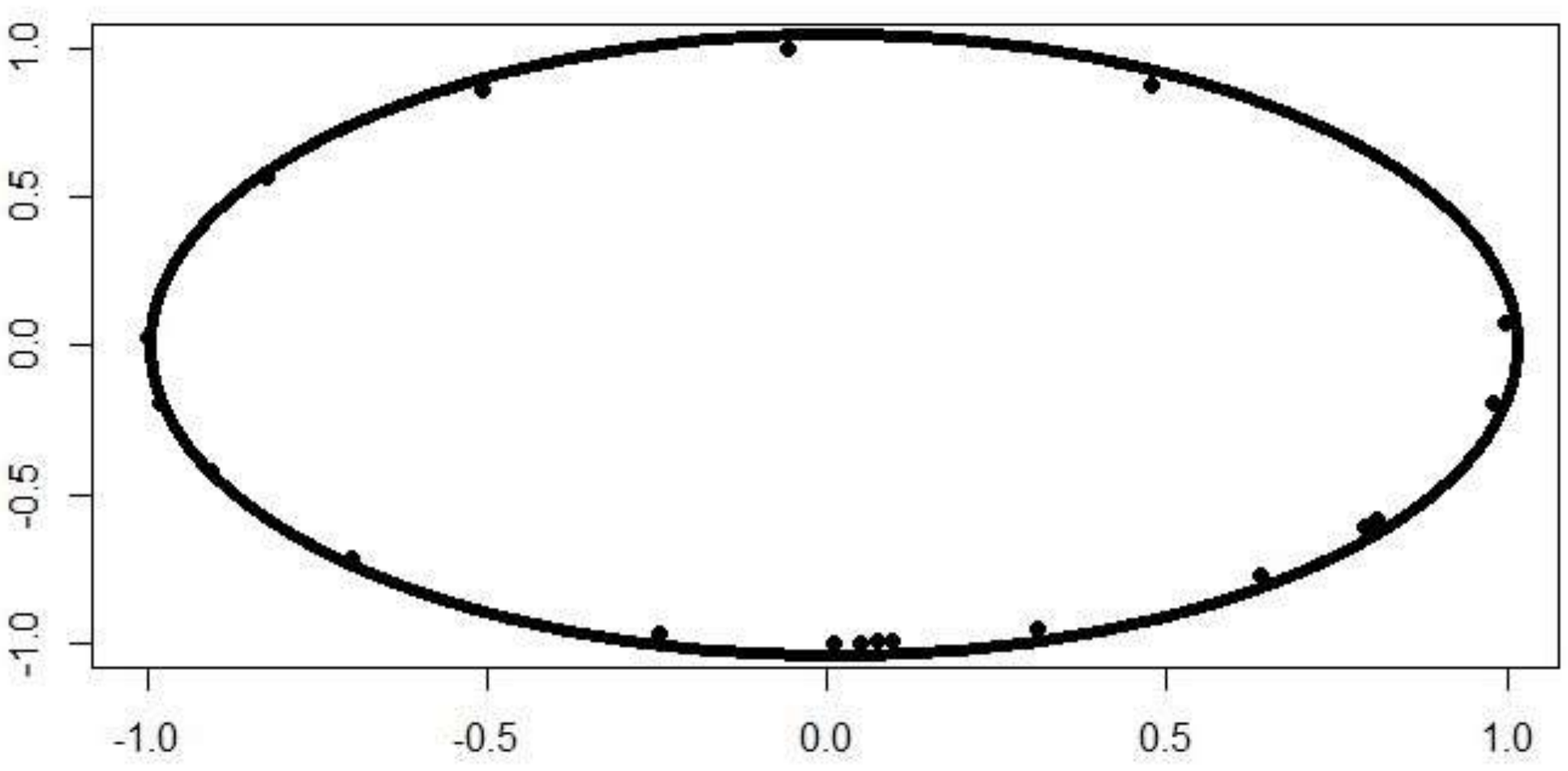}
			\caption{continuous space underlying the data}
		\end{subfigure}
	\caption{Plots of observed data and the underlying continuous space.}
	\label{fig:circle}
	\end{figure}
\end{e1}
Consequently, one needs to define a criterion which specifies the connectivity of points in $\mathds{X}$. We can define a criterion based on the neighbourhood relationship between the data points, and a that can be done by introducing a metric on the set $\mathds{X}$ or some sort of dissimilarity measure for sampled data points. Thus, TDA takes input not in the form of the usual data matrix but in the form of some distance matrix or metric space ($\mathds{X},\delta $), called \textbf{point cloud}, where $\delta$ is any metric such as Euclidean metric.

Now, given a point cloud ($\mathds{X},\delta $) or a distance matrix, to specify the connectivity of points in $\mathds{X}$, one has to choose a threshold, $\epsilon \geq 0$, to decide which points are close or similar to each other. For instance, we consider two points $x_{i} \text{ and } x_{j}$ in $\mathds{X}$ to be connected, or similar, in the sense that $\delta(x_{i},x_{j})\leq \epsilon$. Thus, given a point cloud ($\mathds{X},\delta $) and a threshold $\epsilon$, a natural approximation of the underlying topological space is the cover of $\mathds{X}$, $i.e$, the union of balls $\displaystyle{\bigcup\limits_{x\in\mathds{X}}B_{\epsilon}(x)\subset\mathds{R}^{n}}$, where for any $x\in \mathds{X}$, $B_{\epsilon}(x) = \{v\in\mathds{R}^{n}:\delta(v,x)\leq\epsilon\}$. However, union of balls $\displaystyle{\bigcup\limits_{x\in\mathds{X}}B_{\epsilon}(x)}$ is not useful since it is not algorithmically tractable. Therefore, one can construct an abstract simplicial complex which is homologically equivalent to $\displaystyle{\bigcup\limits_{x\in\mathds{X}}B_{\epsilon}(x)}$. In algebraic topology, a well known theorem, \textbf{The Nerve theorem} (See: \cite{10.3389/frai.2021.667963}), guarantees that homology of $\displaystyle{\bigcup\limits_{x\in\mathds{X}}B_{\epsilon}(x)}$ is same as homology of the $\check{C}$ech complex which is defined as follows:      

\begin{d1}
\textbf{The $\check{C}$ech complex:} Suppose that we are given a point cloud ($\mathds{X},\delta$), where $\mathds{X} = \{x_{1},x_{2}, \ldots,  x_{n} \}$, is a set consist of observed sample points and a threshold $\epsilon\geq0$, then one can construct an abstract simplicial complex, called $\check{C}$ech complex, denoted as $\mathcal{C}(\mathds{X},\epsilon)$, in the following way:
\begin{itemize}
  \item  Elements in $\mathds{X}$ belongs to $\mathcal{C}(\mathds{X},\epsilon)$, that is, $\{x_{i}\}\in \mathcal{C}(\mathds{X},\epsilon)$, for all $i = 1, 2 \ldots n$.
    \item A k-simplex $[x_{i_{0}},x_{i_{1}}\ldots x_{i_{k}}]\in \mathcal{C}(\mathds{X}, \epsilon)$ if $\bigcap\limits_{j=0}^k B_{\epsilon}(x_{i_{j}})\neq\phi$, 
    where $B_{\epsilon}(x)$ is ball of radius $\epsilon$ around the point $x$.
\end{itemize}
\end{d1}

\noindent The $\check{C}$ech complex provides a procedure to construct an abstract simplicial complex from the data but its construction is not computationally tractable. Therefore, one usually constructs the following abstract simplicial complex which is computationally feasible.

\begin{d1}
\textbf{The Vietoris-Rips complex:}
Suppose that we are given a point cloud ($\mathds{X},\delta$), where $\mathds{X} = \{x_{1},x_{2}, \ldots,  x_{n} \}$, is a set consisting of observed sample points and a threshold $\epsilon\geq0$, then one can construct an abstract simplicial complex, called Vietoris-Rips complex or Rips complex, denoted as $\mathcal{R}(\mathds{X},\epsilon)$, in the following way:
\begin{itemize}
  \item Elements in $\mathds{X}$ belongs to $\mathcal{R}(\mathds{X},\epsilon)$, that is, $\{x_{i}\}\in \mathcal{R}(\mathds{X},\epsilon)$, for all $i = 1, 2 \ldots n$.
    
    \item A k-simplex $[x_{i_{0}},x_{i_{1}}\ldots x_{i_{k}}]\in \mathcal{R}(\mathds{X}, \epsilon)$ if $B_{\epsilon}(x_{i_{j}})\bigcap B_{\epsilon}(x_{i_{m}})\neq\phi$ for all $0\leq j < m \leq k$
\end{itemize}
\end{d1}

\noindent The usefulness of the Rips complex is justified by the following nested relationship with the $\check{C}$ech complex which ensures that the Rips complex is an appropriate approximation of the $\check{C}$ech complex to estimate the homology of $\mathcal{X}$:
$$\mathcal{C}(\mathds{X},\epsilon) \subseteq \mathcal{R}(\mathds{X},2\epsilon)\subseteq\mathcal{C}(\mathds{X},2\epsilon).$$

\subsection{Simplicial homology} \label{Simplicial homology}
Here, we shall briefly define the notion of homology for simplicial complexes. Homology counts the number of holes in a topological space. We will require the following notions to define the holes in a topological space.

\begin{d1}

\textbf{Orientation of a simplex:} Given a finite simplicial complex $K$, let $\sigma$ be a p-simplex in $K$ with the vertex set $\{x_{1},\ldots x_{p}\}$, where $p > 0$ be an integer. Define two orderings of its vertex set to be equivalent if they differ from one another by an even permutation, and thus, the ordering of vertices falls into two equivalence classes. Each of these equivalence classes is called an \textbf{orientation} of $\sigma$. We denote an oriented k-simplex with the vertex set $\{x_{1},\ldots x_{p}\}$ as $[x_{1},\ldots x_{p}]$. 

\end{d1}

\begin{d1}
\textbf{Simplicial chains:} Let $K$ be a finite simplicial complex then, for any integer $p \geq 0$, a p-chain on $K$ is defined to be formal linear combinations of p-simplices in K, that is any p-chain can be written as $\sum\limits_{i=1}^{m} \alpha_{i}\sigma_{i}$, where $\sigma_{i}$ is a p-simplex in $K$, $m$ is the number of p-simplices in $K$ and $\alpha_{i}$'s are integers from some field $\mathds{F}$. An empty chain is denoted as 0 and defined as $0 = \sum\limits_{i=1}^{m} 0.\sigma_{i}$.

\noindent In TDA, one usually takes coefficients $\alpha_{i}$'s from the field $\mathds{Z}/2\mathds{Z}$ $:=$ $\mathds{Z}_{2}$ for computational simplicity. Here $\mathds{Z}_{2}$ is a field of elements $\alpha$ modulo 2, where $\alpha\in\mathds{Z}$, $\mathds{Z}$ is the set of integers. We denote $C_{p}(K,\mathds{F})$ to be the space of p-chains over the field $\mathds{F}$. In this article, we take $\mathds{F}$ to be $\mathds{Z}_{2}$. Thus, if m denotes the number of p-simplices in $K$ and $\sigma_{i}$ denotes a p-simplex in $K$, $i = 1,\ldots,m$, then space of p-chains is defined as follows:

$$C_{p}(K,\mathds{F}) := \left\{\gamma = \sum\limits_{i=1}^{m} \alpha_{i}\sigma_{i}: \alpha_{i}\in\mathds{Z}_{2}\right\}.$$

\noindent It can be verified that $C_{p}(K,\mathds{F})$ is a vector space over the field $\mathds{F}$.

\end{d1}

\begin{d1}

\textbf{Boundary homomorphism:} Given a simplicial complex $K$, the $p^{th}$ boundary homomorphism is a linear transformation from $C_{p}(K,\mathds{F})$ to $C_{p-1}(K,\mathds{F})$, also referred as boundary operator. We denote the boundary of a p-simplex $\sigma$ by $\partial_{p}(\sigma)$.

\noindent Let $\sigma = [x_{1}, x_{2},\ldots,x_{p}]$ be an oriented p-simplex with $p > 0$, we define 

$$\partial_{p}(\sigma) = \partial_{p}[x_{1}, x_{2},\ldots,x_{p}] = \sum\limits_{i=0}^{p} (-1)^{i} [x_{1}, \ldots,\hat{x}_{i},\ldots, x_{p}],$$ where $[x_{1}, \ldots,\hat{x}_{i},\ldots, x_{p}]$ is $(p-1)$- face of $\sigma$ obtained by deleting $x_{i}$ from $[x_{1},\ldots,x_{p}]$. Here the symbol $\hat{x}_{i}$ means that we obtain $(p-1)$- face of $\sigma$ by deleting the vertex $x_{i}$ from $[x_{1},\ldots,x_{p}]$.

\noindent Note that for $\mathds{F}$ = $\mathds{Z}_{2}$, boundary of a p-simplex is defined as follows: 

$$\partial_{p}(\sigma) = \partial_{p}[x_{1}, x_{2},\ldots,x_{p}] = \sum\limits_{i=0}^{p} [x_{1}, \ldots,\hat{x}_{i},\ldots, x_{p}].$$

\noindent It can be verified that for any p-simplex $\sigma$, $p\geq1$, we have $\partial_{p-1}\partial_{p}(\sigma) = 0$. This property of boundary operators is also referred as \textbf{fundamental lemma of simplicial homology}, which is essential to define homology. Here, it should be noted that by convention, the boundary operator $\partial_{0}$ maps a 0-simplex to an empty chain 0. Besides, the fundamental lemma of simplicial homology allows us to define an algebraic structure, called chain complex which allows us to compute homology of $K$.
\end{d1}

\begin{d1}

\textbf{Chain complex:} The chain complex of a finite simplicial complex $K$ of dimension n is the sequence of vector spaces $C_{n}(K,\mathds{F})$ connected with the corresponding boundary operators:
$$ 0 \xrightarrow{\partial_{n+1}} C_{n}(K,\mathds{F}) \xrightarrow{\partial_{n}} C_{n-1}(K,\mathds{F}) \xrightarrow{\partial_{n-1}}  \ldots\xrightarrow{\partial_{2}}C_{1}(K,\mathds{F})\xrightarrow{\partial_{1}}C_{0}(K,\mathds{F})\xrightarrow{\partial_{0}} 0.$$
For, $p\in\{0,\ldots n\}$, we define the set of p-cycles as follows:   
$$Z_{p}(K) := \text{ kernel } (\partial_{p}:C_{p}(K,\mathds{F})\xrightarrow{}C_{p-1}(K,\mathds{F})) = \{\gamma\in C_{p}(K,\mathds{F}): \partial_{p}(\gamma) = 0\}.$$
The set of p-boundaries is defined as follows:
$$
 \begin{aligned}
 B_{p}(K) &:= \text{image} (\partial_{p+1}:C_{p+1}(K,\mathds{F})\xrightarrow{}C_{p}(K,\mathds{F})) \\
 &= \{\gamma\in C_{p}(K,\mathds{F}):\gamma = \partial_{p+1}(\gamma^{*}),\text{ for some }\gamma^{*} \in C_{p+1}(K,\mathds{F})\}\\
\end{aligned}
 $$
\end{d1}

\begin{d1}
\textbf{Homology:} Given a finite simplicial complex $K$ and its associated chain complex, it is evident that $B_{p}(K)$ and $Z_{p}(K)$ are subspaces of $C_{p}(K,\mathds{F})$ and according to fundamental lemma of simplicial homology, we have the following: 
$$B_{p}(K)\subset Z_{p}(K)\subset C_{p}(K,\mathds{F}).$$ For any non-negative integer $p$, we denote the $p^{th}$ homology of $K$ as $H_{p}(K)$, which is defined to be the following quotient vector space:
$$H_{p}(K) := Z_{p}(K)/B_{p}(K)$$

\noindent Note that $H_{p}(K)$ is a vector space and its elements are called homology classes of $K$.  
\end{d1}

\begin{d1}
\textbf{Betti numbers:} The $p^{th}$ Betti number $\beta_{p}$ is defined to be the dimension of the vector space $H_{p}(K)$, $i.e$ $\beta_{p} := \text{ dim }(H_{p}(K))$. 
\end{d1}

\noindent Note that the calculation of Betti numbers involves choosing a value $\epsilon > 0$ to construct a simplicial complex $K$, which makes Betti numbers unstable. Suppose that  $K_{\epsilon}$ denotes a simplicial complex constructed at a fixed value of $\epsilon$, then note that the homology of $K_{\epsilon}$ changes as $\epsilon$ changes. Therefore, instead of choosing a single value of $\epsilon$, one tracks the evolution of topological features over all the possible values of $\epsilon$. This idea leads to the notion of persistent homology introduced by \cite{Edelsbrunner(2008)}. 

\begin{d1}
\textbf{Filtrations:} A filtration of a simplicial complex $K$ is a nested family of sub-complexes $(K_{\epsilon})_{\epsilon\in T}$, where $T\subset \mathds{R}$, such that for any $\epsilon, \epsilon^{'} \in T$ we have the following:

\begin{itemize}
    \item $\displaystyle{K = \bigcup\limits_{\epsilon\in T}K_{\epsilon}}$.
    \item $K_{\epsilon}\subseteq K_{\epsilon^{'}}$, for any $\epsilon \leq \epsilon^{'} \in T$. 
\end{itemize}
\end{d1}

\begin{d1}
\textbf{Persistent homology:} For a given filtration $(K_{\epsilon})_{\epsilon\in T}$ of a simplicial complex $K$ and for a non-negative integer $p$, we obtain a sequence of $p^{th}$-homology vector spaces $H_{p}(K_{\epsilon})$, where the inclusions $K_{\epsilon}\subseteq K_{\epsilon^{'}}$, for any $\epsilon \leq \epsilon^{'} \in T$, induce linear maps $f_{p}^{\epsilon, \epsilon^{'}}$ between $H_{p}(K_{\epsilon})$ and $H_{p}(K_{ \epsilon^{'}})$. Such a sequence of vector spaces together with the linear maps connecting them is called the persistence module. Therefore, given a persistence module associated with the simplicial complex $K$, we define the $p^{th}$ persistent homology $H_{p}^{\epsilon, \epsilon^{'}}(K)$ as follows:

$$H_{p}^{\epsilon, \epsilon^{'}}(K) := Z_{p}(K_{\epsilon})/ (B_{p}(K_{\epsilon})\cap Z_{p}(K_{\epsilon^{'}})),$$ where $Z_{p}(K)$ and $B_{p}(K)$ denote the set of p-cycles and the set of p-boundaries of a simplicial complex $K$, respectively. Note that here, $H_{p}^{\epsilon, \epsilon^{'}}(K)$ contains p-dimensional holes of $K_{\epsilon}$ that are still present in $K_{\epsilon^{'}}$. Also, note that $H_{p}^{\epsilon, \epsilon}(K) = H_{p}(K_{\epsilon})$.
\end{d1}

\begin{d1}
\textbf{Persistent Betti numbers:} For any $\epsilon \leq \epsilon^{'}$, the $p^{th}$ persistent Betti number $\beta_{p}^{\epsilon, \epsilon^{'}}$ is defined to be the dimension of the vector space $H_{p}^{\epsilon, \epsilon^{'}}(K)$, $i.e$, $\beta_{p}^{\epsilon, \epsilon^{'}}$ $:=$ dim $(H_{p}^{\epsilon, \epsilon^{'}}(K))$. Note that, for $\epsilon = \epsilon^{'}$, the $p^{th}$ persistent Betti number $\beta_{p}^{\epsilon, \epsilon^{'}}$ coincide with the usual Betti number of $K_{\epsilon}$.  
\end{d1}

\section{Problem Formulation and Main Results}\label{Problem Formulation}

We are given a random sample $X_{1}, \ldots, X_{n}$ from some probability measure $\mathbb{P}$ supported on an unknown compact set $\mathcal{X} \subseteq \mathds{R}^d$, $d\geq 1$, and our goal is to infer the topology of the unknown set $\mathcal{X}$, based on the observed sample. In particular, we are interested in the homology of the set $\mathcal{X}$ based on the observed data. In this study, we assume $\mathcal{X}$ to be a compact manifold (See, Definition 2.5) embedded in $\mathds{R}^d, d \geq 1$.

 Our goal is to perform one sample test for homological equivalence of $\mathcal{X}$ to some hypothesized set $\mathcal{X}_{0}$. Moreover, we shall extend this setup to two sample problems. As Betti numbers characterize the homology of a topological space, we want to test the statistical significance of calculated Betti numbers of $\mathcal{X}$ to the Betti numbers of $\mathcal{X}_{0}$. Further, since $\mathcal{X}$ is known only up to the random sample $X_{1}, X_{2},\ldots, X_{n}$, we approximate the topological space $\mathcal{X}$ by a simplicial complex constructed from the random sample $X_{1}, X_{2},\ldots, X_{n}$. We consider $\check{C}$ech and Vietoris-Rips complex as a simplicial complex constructed from the data to approximate the topological space $\mathcal{X}$.  Specifically,  $\check{C}$ech complex is used for theoretical purposes, and the Vietoris-Rips complex is used for computational purposes to calculate the Betti numbers. 
 
Note that, one needs to choose a positive threshold $\epsilon$ to construct $\check{C}$ech or Rips complex, the threshold $\epsilon$, also referred to as `proximity parameter'. The threshold $\epsilon$ is a tuning parameter, and one can view topological features at different scales by changing the value of $\epsilon$. Therefore, the behavior of the $\check{C}$ech and Rips complex depends on the threshold $\epsilon$. Since we are interested in the asymptotic properties of these random complexes, we choose $\epsilon$ as a function of sample size $n$, i.e.,  $\epsilon \equiv \epsilon \left( n \right ) $ such that $ \epsilon \left( n \right ) \xrightarrow{} 0 \text{ as } n \xrightarrow{} \infty$. Thus, depending on $\epsilon ( n )$, there are three regimes in which the limiting behavior of the random complexes differs significantly. The subcritical regime is when $n\epsilon^{d} \xrightarrow{} 0 \text{ as } n \xrightarrow{} \infty$, the critical regime ( the thermodynamic regime ) is when $n\epsilon^{d} \xrightarrow{} \theta \in \left( 0, \infty\right ) \text{ as } n \xrightarrow{} \infty $ and the supercritical regime is when $n\epsilon^{d} \xrightarrow{} \infty \text{ as } n \xrightarrow{} \infty$, $d \geq 1$ is dimension of the data. We establish consistency of the proposed tests under the critical and the supercritical regimes. 

Now, we shall formulate the testing of the hypothesis problem and state the main theorems, for one and two sample problems, in the following two subsections:  

\subsection{One Sample Test} \label{Subsection:3.2}

Let $X_{1}, \ldots, X_{n}$ be a random sample from a probability measure $\mathbb{P}$, and for all $i = 1, \ldots, n$, $X_{i}$ is supported on $\mathcal{X}\subseteq \mathds{R}^d$, $d\geq 1$. Further, we assume that $\mathcal{X}$ is an unknown compact manifold embedded in $\mathds{R}^d, d \geq 1$. If two sets $\mathbb{X}$ and $\mathbb{Y}$ are homologically equivalent then we denote them as $\mathbb{X}\cong\mathbb{Y}$, and $\mathbb{X}\ncong\mathbb{Y}$ denotes that $\mathbb{X}$ and $\mathbb{Y}$ are not homologically equivalent. Thus, given a set $\mathcal{X}_{0}$, we want to test the following hypothesis: 
\begin{equation} \label{ equation:3.1 }
    H_{0} : \mathcal{X} \cong\mathcal{X}_{0} \text{ vs } H_{1} : \mathcal{X} \ncong\mathcal{X}_{0}
\end{equation}

Now, since Betti numbers characterize the homology of a set, we can reformulate the hypothesis in Equation \ref{ equation:3.1 } in terms of Betti numbers. Suppose that $\mathcal{X}$ and $\mathcal{X}_{0}$ are subsets of d-dimensional Euclidean space, and we denote $\beta_{i}$ and $\beta_{i,0}$ as the $i^{th}$ Betti numbers of $\mathcal{X}$ and $\mathcal{X}_{0}$, respectively, for all $i = 0,1,\ldots, d-1$. Let $\boldsymbol{\beta} = (\beta_{0}, \beta_{1},\ldots,\beta_{d-1})^T$ and $\boldsymbol{\beta_{0}} = (\beta_{0,0}, \beta_{1,0},\ldots,\beta_{d-1,0})^T$, where $\boldsymbol{\beta}$ is unknown, and $\boldsymbol{\beta_{0}}$ is known. One can equivalently formulate the statement of the problem, in terms of $\boldsymbol{\beta}$ and $\boldsymbol{\beta_{0}}$, as follows:
\begin{equation} \label{equation:3.2}
    H_{0}^*: \boldsymbol{\beta} = \boldsymbol{\beta}_{0} \text{ vs } 
H_{1}^*: \boldsymbol{\beta} \neq \boldsymbol{\beta}_{0},
\end{equation}
where $``="$ and $``\neq"$ denote the component-wise equality and inequality, respectively. 

Note that in the assertion of testing of hypothesis problem in Equation \ref{equation:3.2}, $\beta_{i}$'s ($i = 0, \ldots d - 1$) are unknown, and $\beta_{i, 0}$'s ($i = 0, \ldots, d - 1$) are known. Hence, one needs to estimate $\beta_{i}$ for $i = 0, \ldots d - 1$ based on the random sample $X_{1}, X_{2},\ldots X_{n}$ to carry out the testing of hypothesis problem $H_{0}^{*}$ against $H_{1}^{*}$. Let $\hat{\beta}_{0,n}, \hat{\beta}_{1,n}, \ldots,\hat{\beta}_{d-1,n}$ be the estimators of $\beta_{0}, \beta_{1}, \ldots, \beta_{d - 1}$, respectively, where $\hat{\beta}_{i,n}$ ($i = 0, \ldots, d - 1$) are precisely the Betti numbers of the $\check{C}$ech or Rips complex constructed by $X_{1}, X_{2},\ldots X_{n}$, and note that $n$ in the subscript $\hat{\beta}_{i,n}$'s, emphasize the fact that estimated Betti numbers depend on the sample size $n$. Now, to test $H_{0}^{*}$ against $H_{1}^{*}$, one can formulate the test statistic in the following way:  In the view of the testing of the hypothesis problem $H_{0}^{*}$ against $H_{1}^{*}$, the test statistic should be based on the appropriate differences between $\hat{\beta}_{i,n}$ and $\beta_{i, 0}$. In this work, we consider the test statistic as the sum of the absolute difference between $\hat{\beta}_{i,n}$ and $\beta_{i, 0}$ ($i = 0, \ldots, d - 1$), i.e., the test statistic $T_{n}$ is as follows: $$T_{n} = \sum\limits_{i = 0}^{d -1}\left|\hat{\beta}_{i,n} - \beta_{i, 0}\right|.$$

\noindent Observe that for a given data, the larger value of $T_{n}$ indicates the rejection of the null hypothesis $H_{0}^{*}$, i.e., $H_{0}^{*}$ will be rejected when $T_{n} > c$, where $c$ is such that $\mathbb{P}[T_{n} > c|H_{0}^{*}] = \alpha$. Here $\alpha\in (0, 1)$ denotes the level of significance of the test.  It is an appropriate place to mention that one can consider any other distance to formulate the test statistic. 

The following theorem states the consistency of the test under the critical regime.
  
 \begin{t1}\label{Theorem 3.1}
 Let $X_{1}, X_{2},\ldots, X_{n}$ be a random sample of size n from a probability measure $\mathbb{P}$ supported on the unknown compact manifold $\mathcal{X} \subseteq \mathds{R}^{d}, d \geq 1$. Then, under the critical regime, the test, for the hypothesis as in Equation \ref{equation:3.2}, based on $T_{n}$ is consistent, i.e., 
$$\mathds{P}_{H_{1}^{*}} \left[T_{n} > c \right] \xrightarrow{} \text{ 1 as n }\xrightarrow{} \infty,$$   where c is such that $P_{H_{0}^{*}}[T_{n} > c] = \alpha$, and $\alpha \in (0,1)$ is the level of significance of the test.
 \end{t1}

The following theorem states the consistency of the test under the supercritical regime, under the following two assumptions:

\noindent \textbf{Assumptions:} 

\noindent \textbf{(A.1)} The underlying support of the data is a unit volume convex body in $\mathbb{R}^{d}, d \geq 1$, i.e., the set $\mathcal{X} \subseteq \mathbb{R}^{d}$ is a compact and convex set with non-empty interior. 

\noindent \textbf{(A.2)} The $\check{C}$ech or Rips complex at a threshold  $\epsilon \equiv \epsilon(n)$ satisfies the following condition:
$$ \liminf_{n} \mathbb{P} \left(\mathcal{C}(\mathds{X}_{n},\epsilon) \text{ is not connected } \right) > 0$$ or $$ \liminf_{n} \mathbb{P} \left(\mathcal{R}(\mathds{X}_{n},\epsilon) \text{ is not connected } \right) > 0,$$
where $\mathcal{C}(\mathds{X}_{n},\epsilon)$ and $\mathcal{R}(\mathds{X}_{n},\epsilon)$ denotes the $\check{C}$ech and  Rips complex respectively. 

\noindent Note that the $\check{C}$ech or Rips complex is connected at a threshold $\epsilon$ if there exists a path between every pair of vertices.
\begin{t1} \label{Theorem 3.2}
Suppose that $X_{1},X_{2},\ldots,X_{n}$ is a random sample of size n from a uniform distribution supported on the manifold $\mathcal{X} \subseteq \mathds{R}^{d}, d \geq 1$. Then, under the supercritical regime with the Assumptions (A.1) and (A.2), for the hypothesis as in Equation \ref{equation:3.2}, the test based on $T_{n}$ is consistent.
\end{t1}

\subsection{ Two Sample Test }

Suppose that we are given two random samples $\mathds{X}_{n_{1}} = \{ X_{1},X_{2}, \ldots, X_{n_{1}} \}$ and $\mathds{Y}_{n_{2}} = \{ Y_{1}, Y_{2}, \ldots,  Y_{n_{2}} \}$ are drawn independently from the probability measures $\mathbb{P}_{1}$ and $\mathbb{P}_{2}$, supported on the unknown compact manifolds    $\mathcal{X}\subset\mathds{R}^{d}$ and    $\mathcal{Y}\subset\mathds{R}^{d}$, $d \geq 1$, respectively. Let $\hat{\beta}_{i}$ and $\hat{\beta}_{i}^{*}$ denote the $i^{th}$ Betti numbers of $\check{C}$ech or Rips complex constructed from $\mathds{X}_{n_{1}}$ and $\mathds{Y}_{n_{2}}$, respectively, for $i = 0, 1, \ldots, d - 1$, and $\boldsymbol{\beta} = (\beta_{0}, \beta_{1}, \ldots, \beta_{d - 1})^T$ and $\boldsymbol{\beta^{*}} = (\beta_{0}^{*},  \beta_{1}^{*}, \ldots, \beta_{d - 1}^{*})^T$ denote the vector containing the unknown Betti numbers of $\mathcal{X}$ and $\mathcal{Y}$, respectively. We now want to test the homological equivalence of $\mathcal{X}$ and $\mathcal{Y}$, and we formulate the testing of the hypothesis problem as follows:
\begin{equation} \label{equation 3.3}
    H_{0}^{**}: \boldsymbol{\beta} = \boldsymbol{\beta}^{*} \text{ against } 
H_{1}^{**}: \boldsymbol{\beta} \neq \boldsymbol{\beta}^{*}
\end{equation}
 
 In the same spirit of formulation of $T_{n}$ for the one-sample problem, we now propose the following test statistic to test $H_{0}^{**}$ against $H_{1}^{**}$:
$$T_{n_{1},n_{2}} = \sum\limits_{i = 0}^{d -1}\left|\hat{\beta}_{i, n_{1}} - \hat{\beta}_{i, n_{2}}^{*}\right|.$$ 

Observe that the larger value of $T_{n_{1},n_{2}}$ indicates the rejection of the null hypothesis $H_{0}^{**}$, $i.e.$, we reject $H_{0}^{**}$ when $T_{n_{1},n_{2}} > c$, where c is such that $\mathbb{P}[T_{n_{1},n_{2}} > c|H_{0}^{**}] \xrightarrow{} \alpha$ as min $\left( n_{1},  n_{2} \right) \xrightarrow{} \infty$. Here $\alpha\in (0, 1)$ denotes the level of significance of the test. The following theorem states the consistency of the two sample tests based on $T_{n_{1},n_{2}}$, under the critical regime.

\begin{t1} \label{ Theorem 3.3 }
Let $\mathds{X}_{n_{1}} = \{X_{1}, \ldots , X_{n_{1}}\}$ and $\mathds{Y}_{n_{2}} = \{Y_{1}, \ldots, Y_{n_{2}}\}$ are two independent random samples of size $n_{1}$ and $n_{2}$ from the probability measures $\mathbb{P}_{1}$ and $\mathbb{P}_{2}$ which are supported on the manifolds $\mathcal{X} \subseteq \mathds{R}^{d}$ and $\mathcal{Y} \subseteq \mathds{R}^{d}, d \geq 1$, respectively. Further assume that   $n_{1}$ and $n_{2}$ are such that $\displaystyle{\frac{n_{1}}{n_{1}+n_{2}}\xrightarrow{}\lambda}\in(0,1)$ as min$\left(n_{1}, n_{2}\right)$ $\xrightarrow{}\infty$. Then under the critical regime, the test, for the hypothesis as in Equation \ref{equation 3.3}, based on $T_{n_{1},n_{2}}$ is consistent $i.e$,
$$\mathds{P}_{H_{1}^{**}} \left[T_{n_{1},n_{2}} > c \right] \xrightarrow{} 1 \text{ as min}\left(n_{1}, n_{2}\right)\xrightarrow{} \infty,$$ where c is such that $P_{H_{0}^{**}}[T_{n_{1},n_{2}} > c] = \alpha$, and $\alpha \in (0,1)$ is the level of significance of the test. 
\end{t1}

The following theorem states the consistency of the test based on $T_{n_{1},n_{2}}$ under the supercritical regime with the same Assumptions ( A.1 ) and ( A.2 ). 

\begin{t1}\label{Theorem 3.4}
Suppose that we are given two independent samples $\mathds{X}_{n_{1}} = \{X_{1}, \ldots , X_{n_{1}}\}$ and $\mathds{Y}_{n_{2}} = \{Y_{1}, \ldots, Y_{n_{2}}\}$ of size $n_{1}$ and $n_{2}$ from the uniform distribution supported on the manifolds $\mathcal{X} \subseteq \mathds{R}^{d}$ and $\mathcal{Y} \subseteq \mathds{R}^{d}, d \geq 1$, respectively. Moreover, assume that both $\mathcal{X}$ and $\mathcal{Y}$ satisfy the Assumption (A.1) and  $n_{1}$ and $n_{2}$ are such that $\displaystyle{\frac{n_{1}}{n_{1}+n_{2}}\xrightarrow{}\lambda}\in(0,1)$ as min$\left(n_{1}, n_{2}\right)$ $\xrightarrow{}\infty$. Then, under the supercritical regime with the assumption (A.2), the test, for the hypothesis as in Equation \ref{equation 3.3}, based on $T_{n_{1},n_{2}}$ is consistent. 
\end{t1}

\section{Simulation study} \label{Simulated data study}
In this section, we perform a Monte Carlo study to show the consistency of the proposed tests. Moreover, we compare power of two sample tests with Robinson's permutation test (See; \cite{Robinson2017}), the test based on mean persistent landscape functions (See; \cite{BUBENIK201791}) and a permutation test based on persistent diagrams using the Wasserstein distance (See; R software library TDAstats). We will refer to these tests as Robinson's test, landscape test, and permutation test.

Now, observe that to implement our proposed test, one needs to choose threshold $\epsilon$ under the critical and supercritical regime. So we choose the following threshold $\epsilon$, under the critical and supercritical regime :

\begin{enumerate}[(i)]
    \item $\displaystyle{\epsilon = n^{-1/d}}$ ( Satisfy the critical regime )
    \item $\displaystyle{\epsilon = \left(\frac{\log n}{n}\right)^{1/d}}$ ( Satisfy the supercritical regime ),
\end{enumerate}
where $n$ is sample size and $d$ is dimension of the data. Further, note that under the supercritical regime, $\epsilon$ should be such that the Assumption (A.2) (See, Section \ref{Problem Formulation}) holds. Therefore, our choice of $\epsilon$ in (ii) is motivated by the result in \cite{Ganesan} which says that ``for a random sample generated from $\left[0, 1\right]^2$ according to a bounded density $f$, if we choose $\displaystyle{\epsilon(n) = \sqrt{\tau \frac{\log n}{n}}}$ to construct a geometric graph $G$, then for a sufficiently small value of $\tau$ ( a constant that does not depend on n), we have the following " $$ \liminf_{n} \mathbb{P} \left( G \text{ is not connected }\right) > 0.$$

Now, in view of the fact that connectivity properties of geometric graphs, $\check{C}$ech complex and Rips complex are the same, one can use the result in \cite{Ganesan} to choose $\epsilon$ such that the Assumption (A.2) holds. However, note that the result in \cite{Ganesan} is given when the data is supported on a set $S = \left[0, 1\right]^2$ and as of now, we are not aware of any generalizations of this result when $S \subset \left[0, 1\right]^d, d \geq 2$. Therefore, while using the test under the supercritical regime, we verify empirically that for the choice of $\epsilon$ as in (ii) the Rips complex satisfies the Assumption (A.2). In other words, under the supercritical regime, we verify that for large $n$, the Rips complex is not connected at a threshold $\epsilon$. 

Furthermore, since, data scaling does not change its topology, we scale the data by dividing each data point by its Euclidean norm. We apply our testing procedure for the data points $\displaystyle{\frac{X_{i}}{||X_{i}||}}$ for all $i = 1,\ldots n$, where $||.||$ denotes the Euclidean norm. 

The following algorithms give the simulated power of the proposed one and two-sample test for the scaled data. Note that in both the algorithms, $1_{A}$ denotes the indicator function on the set $A$ and in the Algorithm \ref{ Algo:1}, $\boldsymbol{\beta_{0}}$ denotes the vector of hypothesised Betti numbers. Next, we provide a simulated data study for the one and two-sample test under the critical and the supercritical regime.

\begin{algorithm}[!ht] 
\SetKwInOut{Input}{Input}\SetKwInOut{Output}{Output}
\caption{Power of one sample test} \label{ Algo:1} 
\Input{A random sample $X_{1},\ldots,X_{n}$, hypothesised Betti numbers $\boldsymbol{\beta_{0}}$, a threshold $\epsilon$, number of replications $r$, level of significance $\alpha$ }
\Output{ Estimated power of the test}

\For{$i = 1$ \KwTo $r$} { \emph{Scale data points by its Euclidean norm }\;
\emph{Compute $T_{n}$ under $H^{*}_{0}$ and $H^{*}_{1}$ for the scaled data}  

}
Estimate critical value $c$ by $\displaystyle{\hat{c} =\left(1- \frac{\alpha}{2}\right)^{th}}$ quantile of $r$ values of $T_{n}$ under $H^{*}_{0}$

Estimated power = $\displaystyle{\frac{\sum\limits_{i = 1}^{r}1_{\displaystyle{\{ T^{'}_{i,n} > \hat{c} \}}}}{r}}$, $T^{'}_{i,n}$ is the $i^{th}$ value of $T_{n}$ under $H^{*}_{1}$

\end{algorithm}

\begin{algorithm}[ht] 
\SetKwInOut{Input}{Input}\SetKwInOut{Output}{Output}
\caption{Power of two-sample test} \label{ Algo:2} 
\Input{Two independent random samples $\{X_{1},\ldots,X_{n}\}$ and $\{Y_{1},\ldots,Y_{n}\}$, a threshold $\epsilon$, number of replications $r$, level of significance $\alpha$ }
\Output{ Estimated power of the test}

\For{$i = 1$ \KwTo $r$} { \emph{Scale data points of both the samples by their Euclidean norms }\;
\emph{Compute $T_{n_{1}, n_{2} }$ for the scaled data under $H^{**}_{0}$ and $H^{**}_{1}$ }  

}
Estimate the critical value $c$ by $\displaystyle{\hat{c} =\left(1- \frac{\alpha}{2}\right)^{th}}$ quantile of $r$ values of $T_{n_{1}, n_{2}}$ under $H^{**}_{0}$

Estimated power = $\displaystyle{\frac{\sum\limits_{i = 1}^{r}1_{\{ \displaystyle{T^{'}_{i,n_{1},n_{2}} > \hat{c}} \}}}{r}}$, $T^{'}_{i,n_{1},n_{2}}$ is the $i^{th}$ value of $T^{'}_{i,n_{1},n_{2}}$ under $H^{**}_{1}$

\end{algorithm}

\subsection{The critical regime} \label{The critical regime} 

In this subsection we provide a simulated data study for one and two sample test under the critical regime. We generate a random sample from a mixture of two von Mises distributions with mixture probabilities 1/3 and 2/3, mean direction parameters $\left(1, 0\right )^T$, $\left(0, 1\right )^T$ and concentration parameters 3, 4. Then we apply one sample homological equivalence testing procedure for this simulated data. Further, we provide a simulation study for two-sample test for the support of a mixture of three von Mises distributions in dimension 3 with mixture probabilities 1/3 for each variate, mean direction parameters $\left(1, 0, 0\right )^T$, $\left(0, 1, 0\right )^T$, $\left(0, 0, 1\right )^T$ and concentration parameter 3, 4, 5, vs the support of 3-D normal distribution with mean vector $\left(0, 0, 0\right )^T$ and covariance matrix $\begin{pmatrix} 
1 & 0.5 & 0.5\\
0.5 & 1 & 0.5\\
0.5 & 0.5 & 1
\end{pmatrix}$.

\noindent In addition, we apply the two-sample testing procedure for a 3-D unit sphere vs a unit cube. Moreover, we compare power of two sample tests with Robinson's test, landscape test, and permutation test. We have calculated power of these tests for the homology in dimension 1 and maximum threshold 4 for the rips filtration. We have applied these tests for 10 replications and the number of permutations is 30. For Robinson's test, we have used the joint loss function for the $q^{th}$ power of the $p^{th}$ Wasserstein distance, where $p = q =1$  and for the landscape test we have calculated the mean of first landscape function over 1000 grid points. 

\subsubsection{Circle} 

\begin{figure}[!ht]
		\centering
		\begin{subfigure}[!ht]{0.45\linewidth}
			\includegraphics[width=\linewidth, height = 3in]{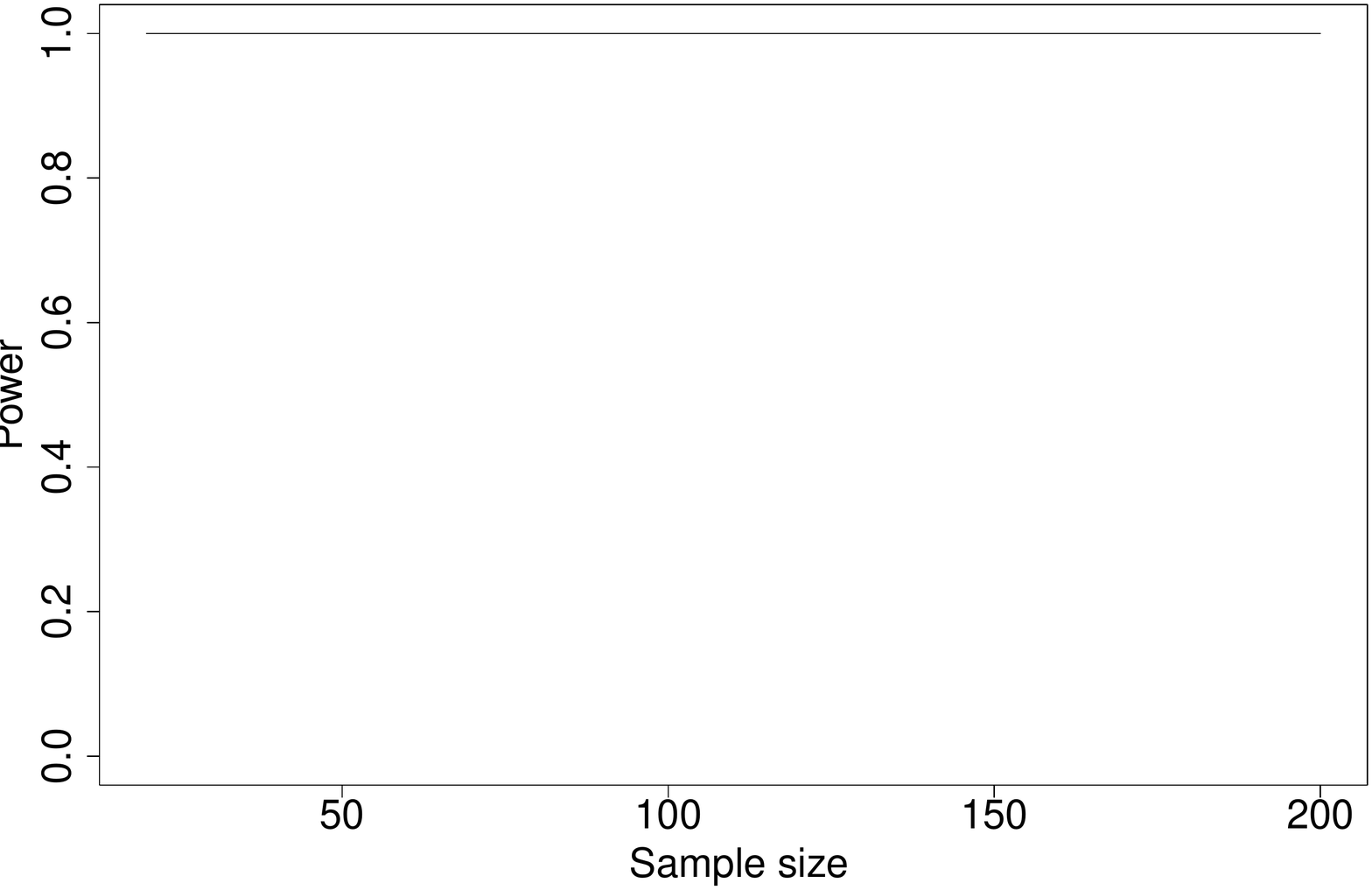}
			\caption{ Circle vs Disk}
		\end{subfigure}
	\hfill
		\begin{subfigure}[!ht]{0.45\linewidth}
			\includegraphics[width=\linewidth, height = 3in]{Powerplot.pdf} 
			\caption{Circle vs support of the bivariate normal}
		\end{subfigure}
	\caption{ Power plots for Circle.}	
	\label{fig:4.1}
	\end{figure}

We generate a random sample of size $n$ from the von Mises distribution and apply the Algorithm \ref{ Algo:1} for $\boldsymbol{\beta_{0}} = \left(1, 1\right)^T, \epsilon = n^{-1/d}, r = 10 \text{ and }\alpha = 0.05$. Then we increase the sample size $n$ to examine the consistency of the test. Note that here $d = 2$ for the circle and we take $n\in\{20, 50, 100, 150, 200\}$. The plots in Figure \ref{fig:4.1} suggest that the test is consistent under the two alternative support unit disk and support of the bivariate normal distribution with mean vector $\left(0, 0\right )^T$ and covariance matrix $\begin{pmatrix}
1 & 0.5\\
0.5 & 1 
\end{pmatrix}$. 

\subsubsection{3-D von Mises vs 3-D Multivariate normal }
\begin{figure}[!ht]
		\centering
		\begin{subfigure}[!ht]{0.45\linewidth}
			\includegraphics[width=\linewidth, height = 3in]{Powerplot.pdf}
			\caption{ Test based on $T_{n_{1},n_{2}}$}
		\end{subfigure}
	\hfill
		\begin{subfigure}[!htt]{0.45\linewidth}
			\includegraphics[width=\linewidth, height = 3in]{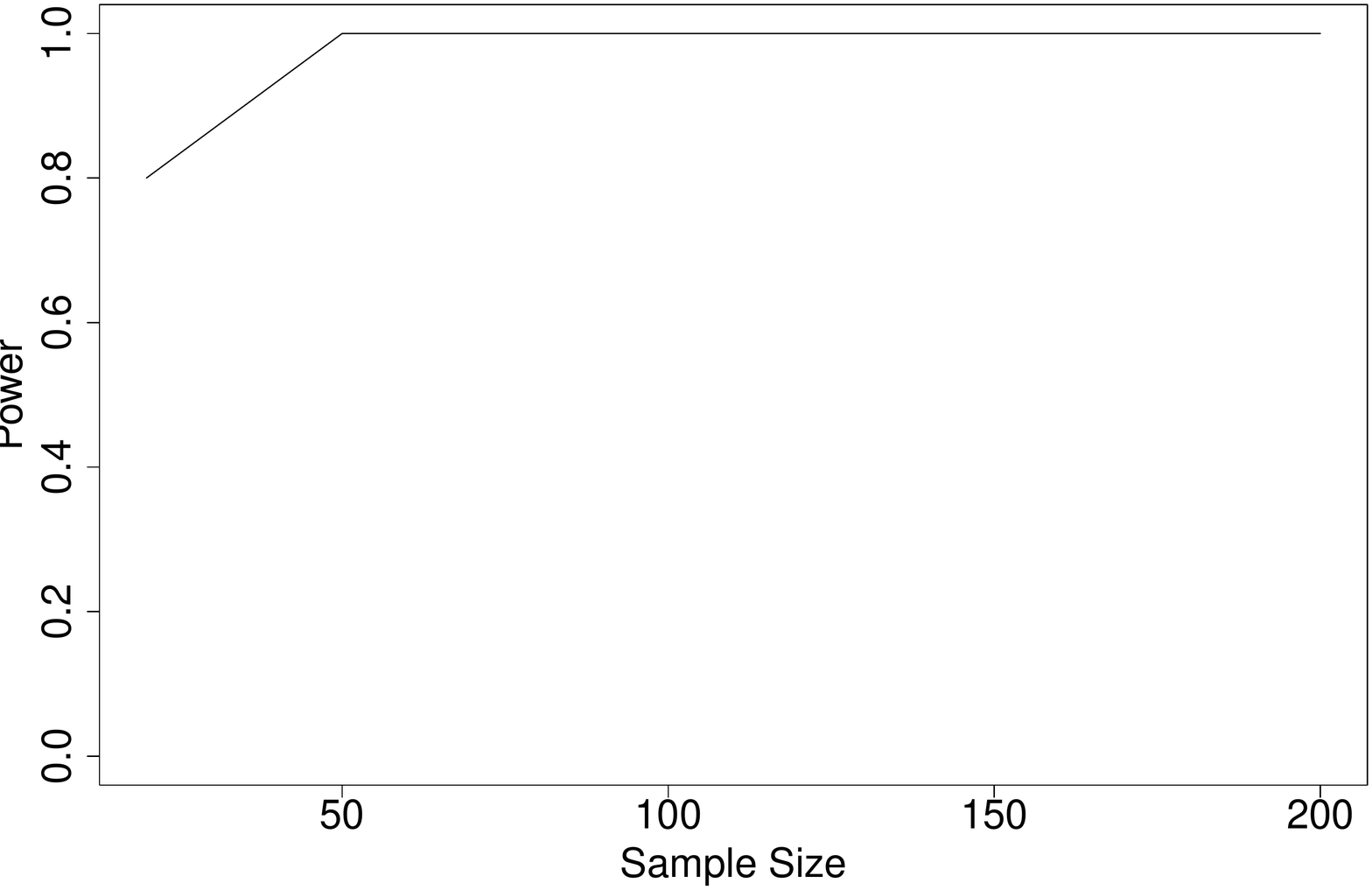}
			\caption{Robinson's test}
		\end{subfigure}
 
             \begin{subfigure}[!ht]{0.45\linewidth}
			\includegraphics[width=\linewidth, height = 3in]{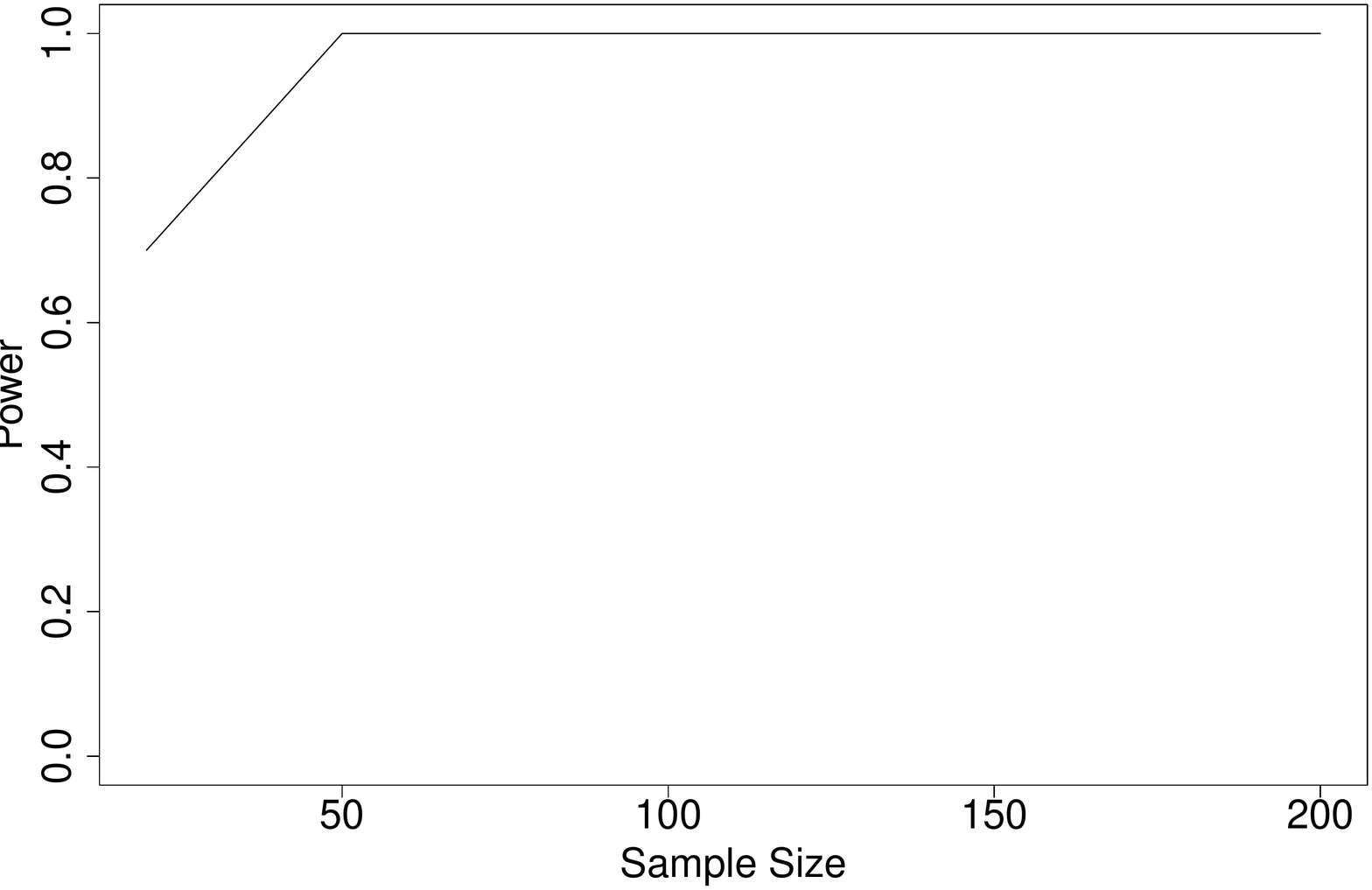}
			\caption{ Landscape test}
		\end{subfigure}
	\hfill
		\begin{subfigure}[!htt]{0.45\linewidth}
			\includegraphics[width=\linewidth, height = 3in]{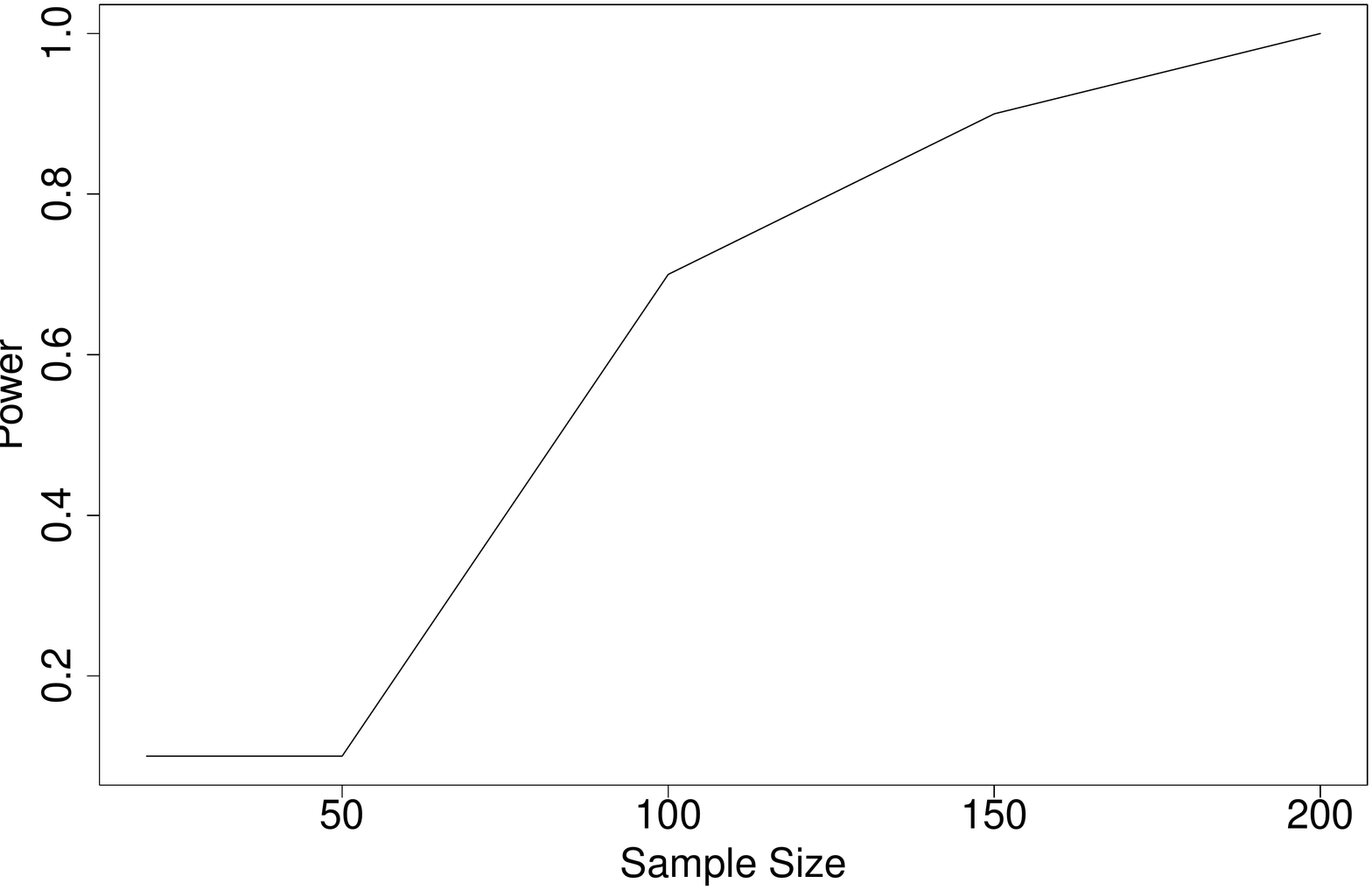}
			\caption{Permutation test} 
		\end{subfigure}
	\caption{von Mises vs Multivariate normal. }	
	\label{fig:4.2}
	\end{figure}

We generate two random samples of the same size $n$ from the 3-D von Mises and 3-D multivariate normal distribution and apply the Algorithm \ref{ Algo:2} for $\epsilon = n^{-1/d}, r = 10 \text{ and }\alpha = 0.05$. Then we increase the sample size $n$ to examine the consistency of the test. Note that here $d = 3$ and we take $n\in\{20, 50, 100, 150, 200\}$. The plots in Figure \ref{fig:4.2} suggest that the test is consistent and performs better than the other tests.

\subsubsection{3-D Sphere vs Unit cube}
\begin{figure}[!ht]
		\centering
		\begin{subfigure}[!htt]{0.45\linewidth}
			\includegraphics[width=\linewidth, height = 3in]{Powerplot.pdf}
			\caption{Test based on $T_{n_{1},n_{2}}$}
		\end{subfigure}
           \hfill
            \begin{subfigure}[!ht]{0.45\linewidth}
			\includegraphics[width=\linewidth, height = 3in]{Powerplot.pdf}
			\caption{ Robinson's test}
		\end{subfigure}
	\begin{subfigure}[!htt]{0.45\linewidth}
			\includegraphics[width=\linewidth, height = 3in]{Powerplot.pdf}
			\caption{Landscape test}
		\end{subfigure}
              \hfill
                \begin{subfigure}[!ht]{0.45\linewidth}
			\includegraphics[width=\linewidth, height = 3in]{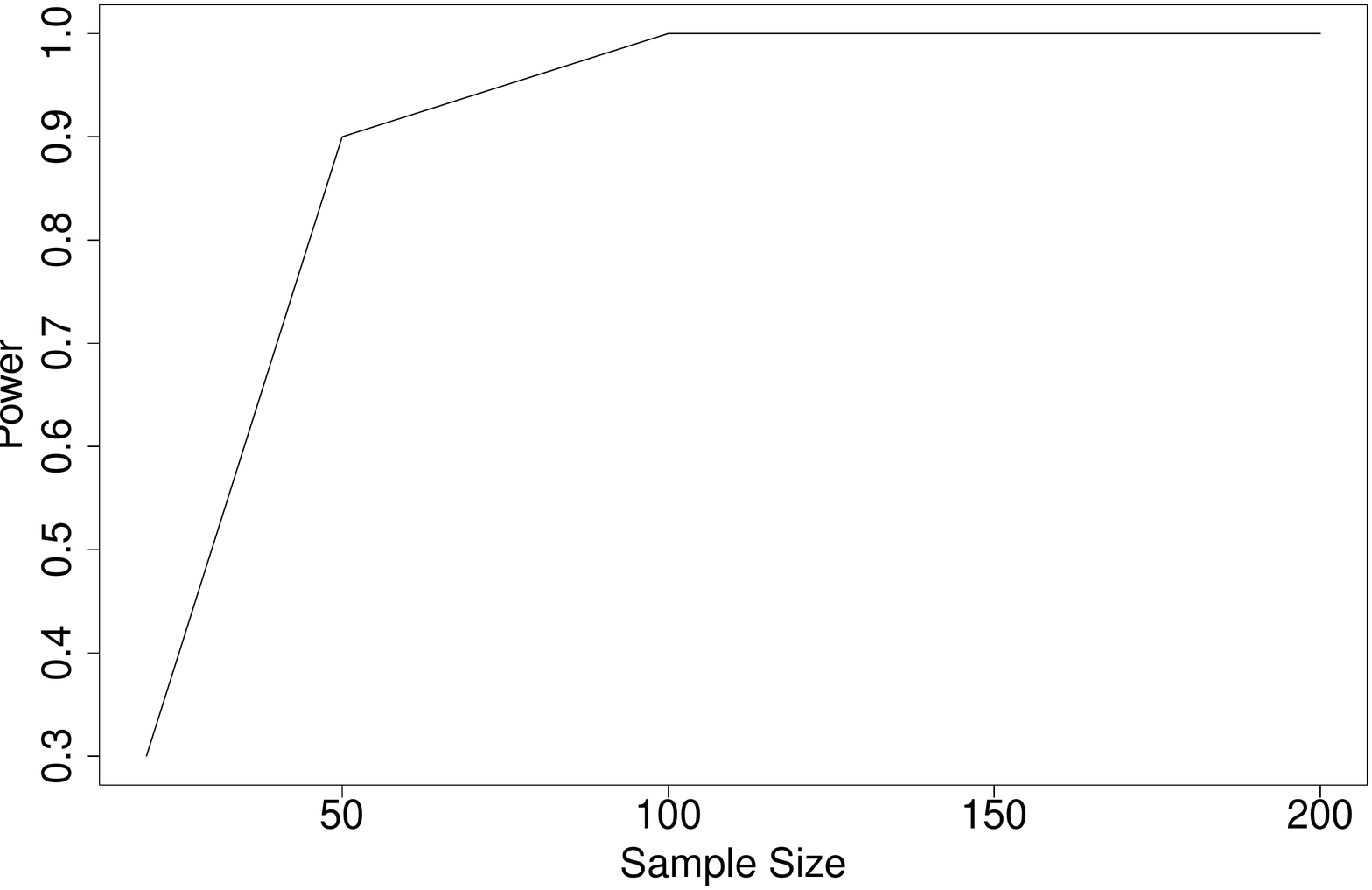}
			\caption{Permutation test }
		\end{subfigure}
	\caption{Sphere vs unit cube.}	
	\label{fig:4.3}
	\end{figure}

We generate two random samples of the same size $n$ from the 3-D Sphere and unit cube according to uniform distribution and apply the Algorithm \ref{ Algo:2} for $\epsilon = n^{-1/d}, r = 10 \text{ and }\alpha = 0.05$. Then we increase the sample size $n$ to examine the consistency of the test. Note that here $d = 3$ and we take $n\in\{20, 50, 100, 150, 200\}$. The plots in Figure \ref{fig:4.3} suggest that the test is consistent and performs equally well in comparison to Robinson's test and landscape test and performs better than the permutation test.  

 \subsection{The supercritical regime} \label{ The supercritical regime}

In this subsection, we provide a simulated data study for one and two sample tests under the supercritical regime. We generate a random sample from the unit disk according to the uniform distribution and apply one sample homological equivalence testing procedure. Further, we provide a simulation study for two-sample test for 3-D sphere vs support of Swiss roll data set (See, R library tdaunif) and 3-D torus vs 3-D sphere. Moreover, we compare power of two sample tests with Robinson's test, landscape test, and permutation test.

We have generated random samples from 3-D torus with the following parametric equation: $$\mathcal{X}_{0}= \{ ((R + r\cos{\theta} \cos{\psi}), (R + r\cos{\theta} \sin{\psi}), r\sin{\theta}) : 0 \leq \theta, \psi \leq 2\pi, 0 < r < R < \infty\}$$

\noindent where $R$ is the distance from the center of the tube to the center of the torus and $r$ is the radius of the tube. We have taken $R$ = 2 and $r$ = 1.

Moreover, we compare power of two sample tests with Robinson's test, landscape test, and permutation test. We have calculated power of these tests for the homology in dimension 1 and maximum threshold 4 for the rips filtration. We have applied these tests for 10 replications and the number of permutations is 30. For Robinson's test, we have used the joint loss function for the $q^{th}$ power of the $p^{th}$ Wasserstein distance with $p = q =1$, and for the landscape test we have calculated the mean of first landscape function over 1000 grid points.

\subsubsection{Unit disk}
\begin{figure}[!ht]
		\centering
		\begin{subfigure}[!ht]{0.45\linewidth}
			\includegraphics[width=\linewidth, height = 3in]{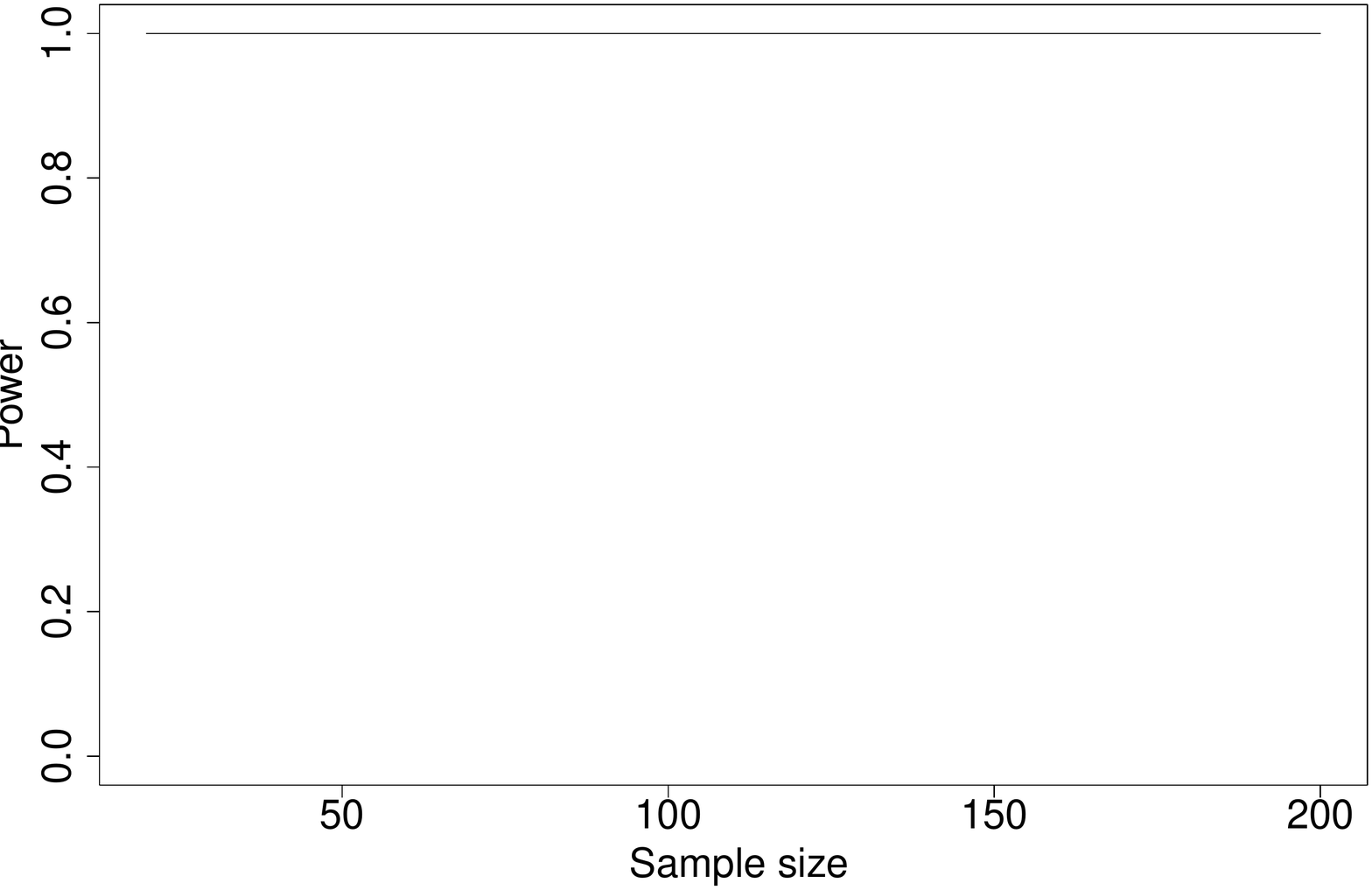}
			\caption{ Disk vs Spiral data} 
		\end{subfigure}
	\hfill
		\begin{subfigure}[!ht]{0.45\linewidth}
			\includegraphics[width=\linewidth, height = 3in]{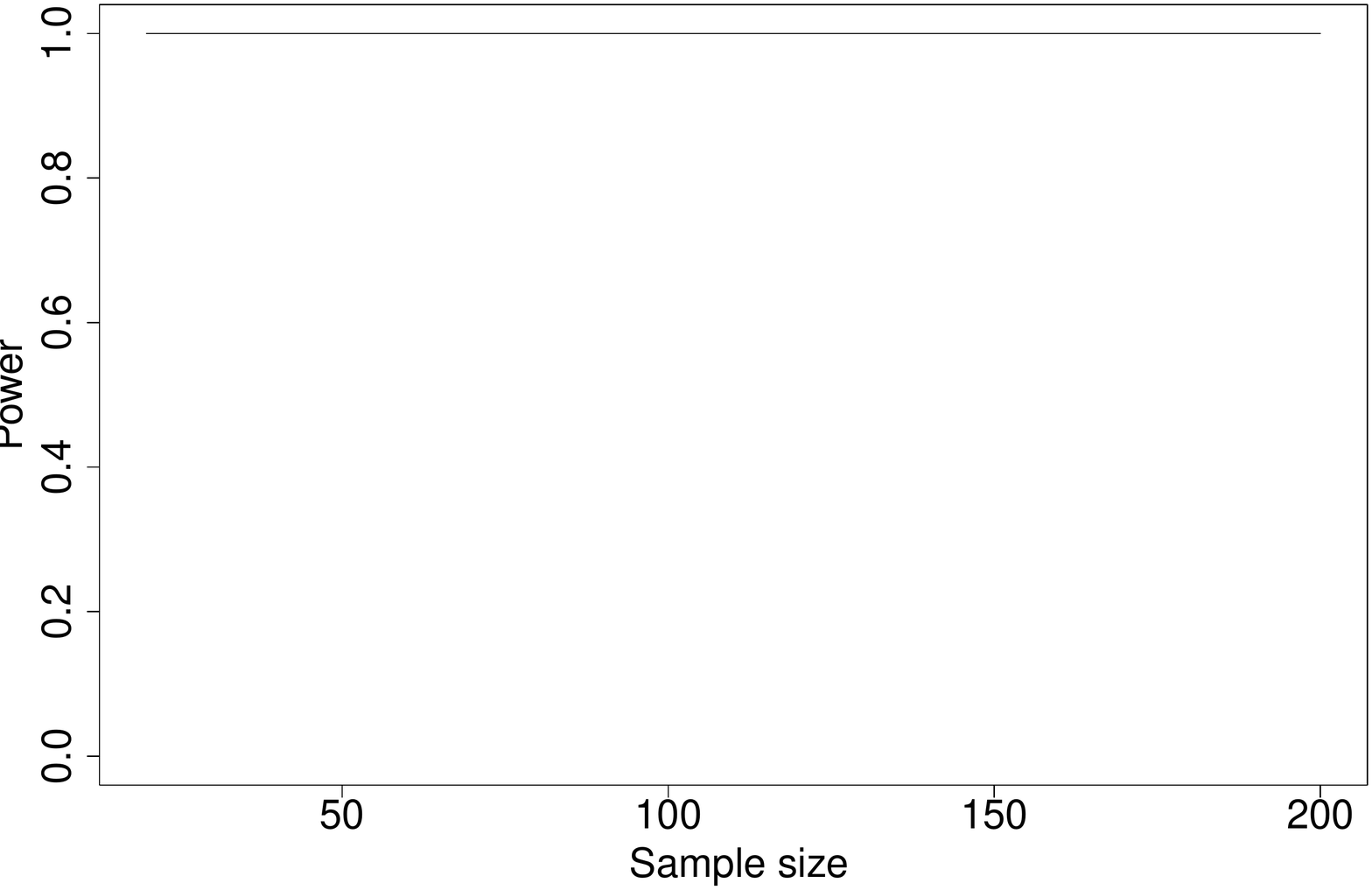}
			\caption{Disk vs Unit square}
		\end{subfigure}
	\caption{Power plots for the Unit disk.}	
	\label{fig:4.4}
	\end{figure}

We generate a random sample of size $n$ from unit disk according to the uniform distribution and apply the Algorithm \ref{ Algo:1} for $\boldsymbol{\beta_{0}} = \left(1, 0\right)^T, \displaystyle{\epsilon = \left(\frac{\log n}{n}\right)^{1/d}}, r = 10 \text{ and }\alpha = 0.05$. Then we increase the sample size $n$ to examine the consistency of the test. Note that here $d = 2$ for the circle and we take $n\in\{20, 50, 100, 150, 200\}$. The plots in Figure \ref{fig:4.4} suggest that the test is consistent under the two alternative support, support of spiral data (See, R library tdaunif) and unit square.  

\subsubsection{3-D Sphere vs Swiss roll data}
\begin{figure}[!ht]
		\centering
                \begin{subfigure}[!htt]{0.45\linewidth}
			\includegraphics[width=\linewidth, height = 3in]{Powerplot.pdf}
			\caption{Test based on $T_{n_{1},n_{2}}$}
		\end{subfigure}
         \hfill
		\begin{subfigure}[!ht]{0.45\linewidth}
			\includegraphics[width=\linewidth, height = 3in]{Powerplot.pdf}
			\caption{Robinson's test}
		\end{subfigure}
	\begin{subfigure}[!htt]{0.45\linewidth}
			\includegraphics[width=\linewidth, height = 3in]{Powerplot.pdf}
			\caption{Landscape test}
		\end{subfigure}
			\hfill
		\begin{subfigure}[!ht]{0.45\linewidth}
			\includegraphics[width=\linewidth, height = 3in]{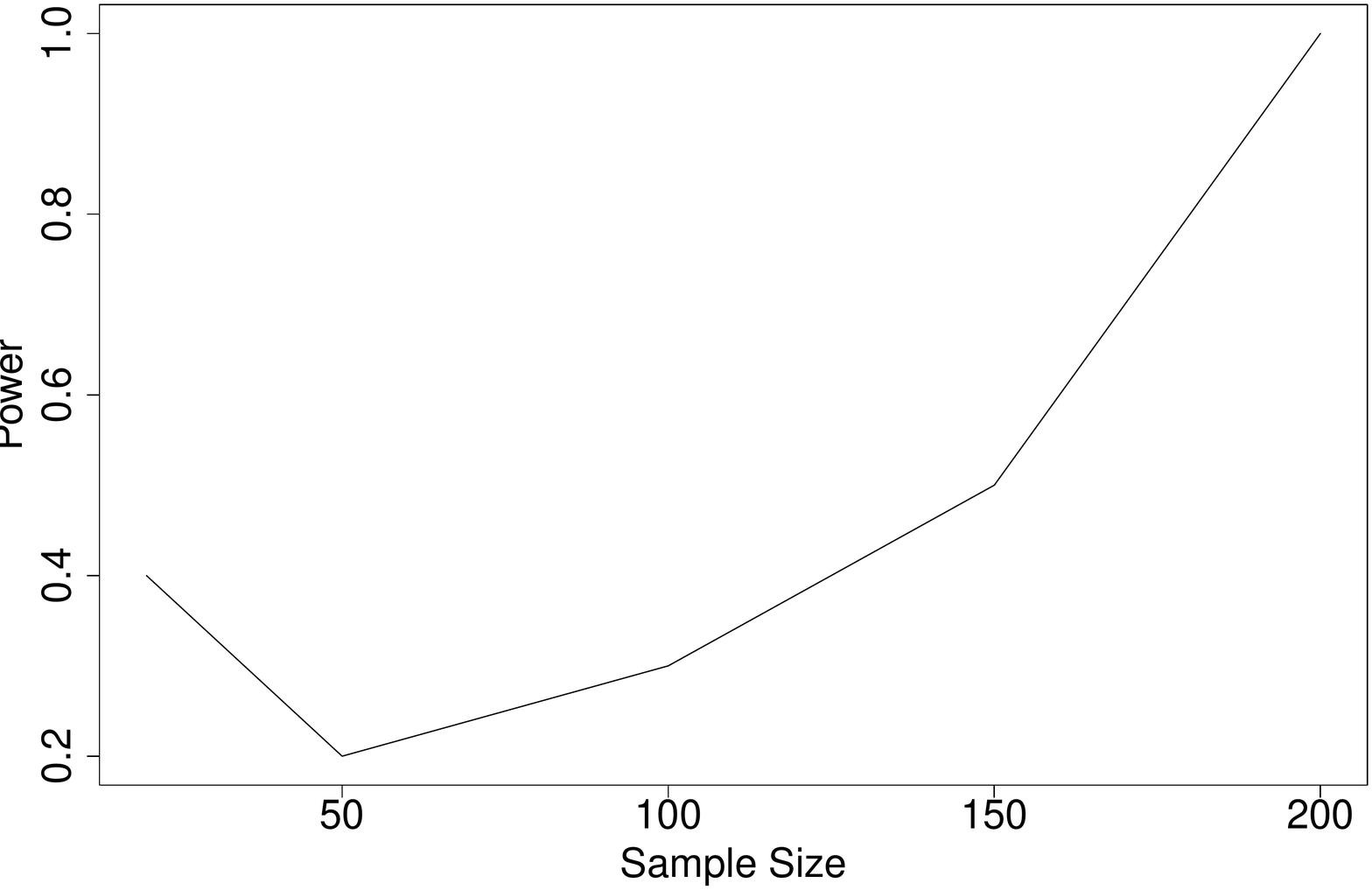}
			\caption{ Permutation test}
		\end{subfigure}
	\caption{Sphere vs Swiss roll data }	
	\label{fig:4.5}
	\end{figure}

We generate two random samples of the same size $n$ from the 3-D sphere and Swiss roll data and apply the Algorithm \ref{ Algo:2} for $\displaystyle{\epsilon = \left(\frac{\log n}{n}\right)^{1/d}}, r = 10 \text{ and }\alpha = 0.05$. Then we increase the sample size $n$ to examine the consistency of the test. Note that here $d = 3$ and we take $n\in\{20, 50, 100, 150, 200\}$. The plots in Figure \ref{fig:4.5} suggest that the test is consistent and performs equally well in comparison to Robinson's test and landscape test and performs better than the permutation test.

\subsubsection{3-D Torus vs 3-D Sphere} 
\begin{figure}[!ht]
		\centering
		\begin{subfigure}[!ht]{0.45\linewidth}
			\includegraphics[width=\linewidth, height = 3in]{Powerplot.pdf}
			\caption{ based on $T_{n_{1},n_{1}}$ }
		\end{subfigure}
	\hfill
	\begin{subfigure}[!htt]{0.45\linewidth}
			\includegraphics[width=\linewidth, height = 3in]{Powerplot.pdf}
			\caption{Robinson's test}
		\end{subfigure}
			\begin{subfigure}[!htt]{0.45\linewidth}
			\includegraphics[width=\linewidth, height = 3in]{Powerplot.pdf}
			\caption{Landscape test}
		\end{subfigure}
		\hfill
		\begin{subfigure}[!ht]{0.45\linewidth}
			\includegraphics[width=\linewidth, height = 3in]{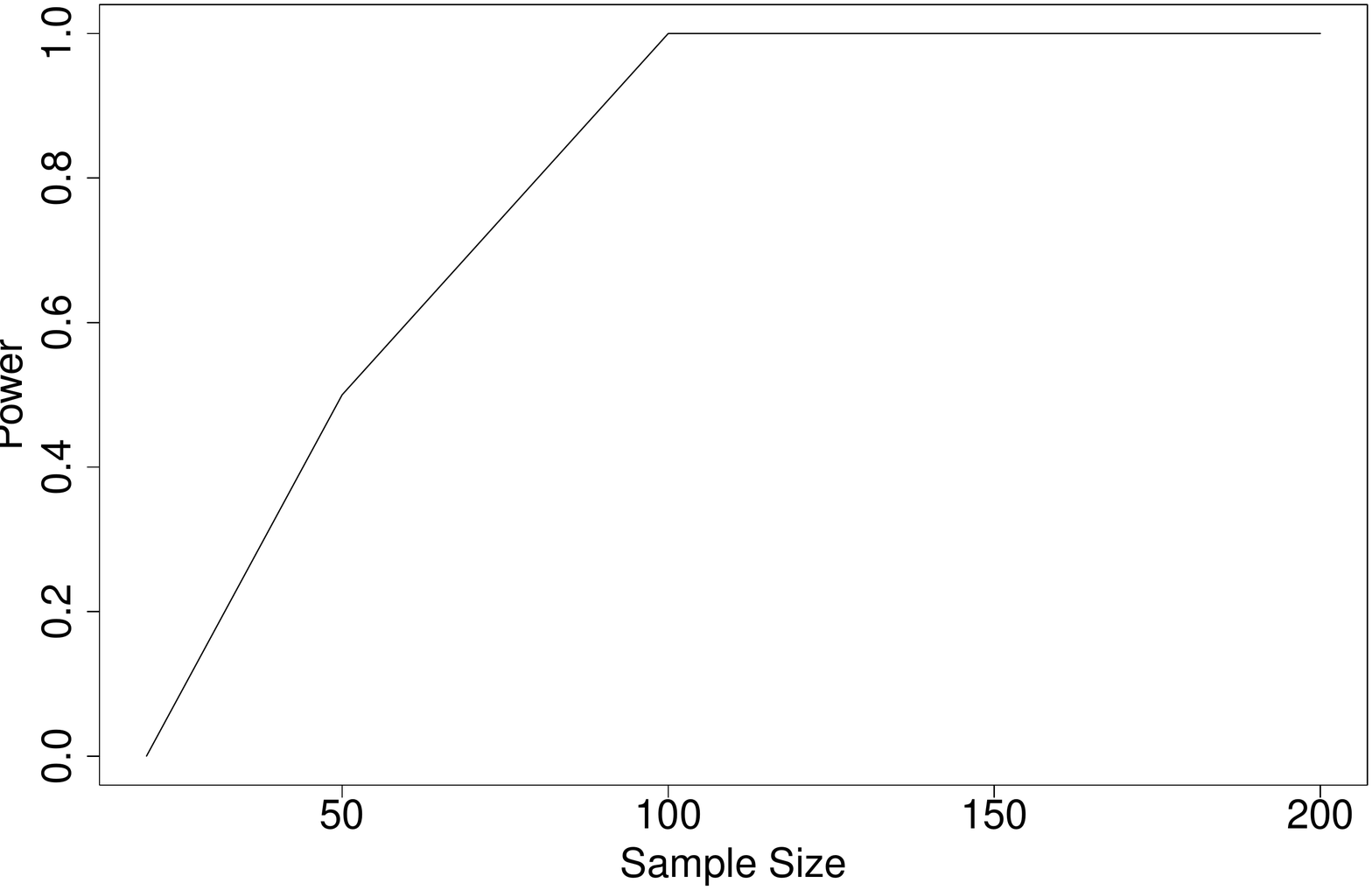}
			\caption{Permutation test }
		\end{subfigure}
	
	\caption{3-D Torus vs 3-D Sphere }	
	\label{fig:4.6}
	\end{figure}

We generate two random samples of the same size $n$ from the 3-D torus and 3-D sphere distribution and apply the Algorithm \ref{ Algo:2} for $\displaystyle{\epsilon = \left(\frac{\log n}{n}\right)^{1/d}}, r = 10 \text{ and }\alpha = 0.05$. Then we increase the sample size $n$ to examine the consistency of the test. Note that here $d = 3$ and we take $n\in\{20, 50, 100, 150, 200\}$. The plots in Figure \ref{fig:4.6} suggest that the test is consistent and performs equally well in comparison to Robinson's test and landscape test, and performs better than the permutation test. 

\section{Conclusion}\label{Conclusion} 
In this article, we have proposed a statistical test based on the Betti numbers to investigate the homological equivalence of the support of distribution-generating data. To the best of our knowledge, this is the first attempt at using Betti numbers to test homological equivalence. Testing the homological equivalence of two spaces has various applications in topological data analysis and plays an important role in manifold learning. Moreover, testing the homological equivalence of two spaces helps in postulating a suitable statistical model for the data. As of now, two sample testing procedures based on persistent homology have been developed by permutating the values of the test statistics under the null hypothesis, and unlike our test based on Betti numbers, most of them use some topological summary, such as persistent diagrams, persistent landscape functions, Betti functions, etc, as a test statistic and applies the permutation test for the null hypothesis.

In general, tests based on persistent homology can be applied for any population or process but are less efficient when the support of the data distribution is not compact as can be seen in Fig: \ref{fig:4.2}. In addition, the proposed test is computationally less expensive, and hence, it is suitable for high-dimensional data and for a large number of sample points. Moreover, the proposed one-sample test has various applications, for instance, tree structure plays a crucial role in modeling an evolutionary phenomenon, thus one can construct a graph from the data and test if it has a tree structure or not. Note that a graph does not have a tree structure if it has loops. Therefore, one can use the proposed one sample test to check if the population $1^{st}$ Betti number $\beta_{1}$ is non-zero or not. For future considerations, we would like to examine these results for other geometric complexes such as the Morse complex, and for the data supported on manifolds that are not embedded in the Euclidean spaces. 

\section{Appendix}\label{Appendix}

In this section, we provide proof of the theorems stated in Section \ref{Problem Formulation}. We state some results from \cite{Bobrowski(2018)} in the following lemmas that will be used in the proof.

\begin{l1} \label{Lemma 1}
(Section 3.2, \cite{Bobrowski(2018)}) Suppose that $\hat\beta_{0,n}$ denotes the $0^{th}$ Betti number of $\check{C}$ech or Rips complex. Then under the critical regime, for some constant $ 0< \xi < 1$, we have the following:
$$ \frac{\hat\beta_{0,n}}{n} \xrightarrow{ a.s } \xi$$
\end{l1}
 
 \begin{l1} \label{Lemma 2}
 (Theorem 4.6, \cite{Bobrowski(2018)}) Suppose that $\hat\beta_{i,n}$ denotes the $i^{th}$ Betti number of $\check{C}$ech or Rips complex and let dimension of the data is $d \geq 2$, then under the critical regime there exists positive constants $n_{0}, \eta_{1}$ and $\eta_{2}$ such that $\eta_{1} < \eta_{2}$, then for all $n > n_{0}$, we have
 $$\eta_{1} < \mathbb{E} \left(\frac{\hat\beta_{i,n}}{n}\right) < \eta_{2}, \text{ for all } i = 0, \ldots d-1.$$ 
 \end{l1}
 
 \begin{l1} \label{Lemma 3}
  (Theorem 4.7, \cite{Bobrowski(2018)}) Suppose that $\hat\beta_{i,n}$ denotes the $i^{th}$ Betti number of $\check{C}$ech or Rips complex, then under the critical regime, for all $i = 1,\ldots d-1, d \geq 2$, we have 
  
  $$ \frac{\hat\beta_{i,n} - \mathbb{E} \left(\hat\beta_{i,n}\right)}{n}\xrightarrow{a.s} 0  .$$
 \end{l1}

\begin{l1}\label{Lemma 4}
(Section 3.3, \cite{Bobrowski(2018)}) Suppose that $\hat\beta_{0,n}$ denotes the $0^{th}$ Betti number of $\check{C}$ech or Rips complex. Then, under the supercritical regime for the choice of $\epsilon$ satisfying the Assumption (A.2),  (Section \ref{Problem Formulation}), $\displaystyle{\frac{\hat{\beta}_{0,n}}{n}}$ is asymptotically almost surely bounded, i.e., there exist constants $0<k<l<\infty$ such that:
$$\mathds{P}\left[k\leq\frac{\hat\beta_{0,n}}{n}\leq l\right]\xrightarrow{}\text{ 1 as n } \xrightarrow{} \infty.$$
\end{l1}
 
\begin{l1} \label{Lemma 5}

 (Theorem 4.9, \cite{Bobrowski(2018)}) Suppose that $\hat\beta_{i,n}$ denotes the $i^{th}$ Betti number of $\check{C}$ech or Rips complex, generated by a uniform distribution on a unit-volume convex body in $\mathds{R}^d, d \geq 2$. Then, under the supercritical regime
with the Assumption (A.1), we have  $\mathds{E}\left(\hat\beta_{i,n}\right) = o(n)$ for all $  i = 1,\ldots d-1 $.

\end{l1}

\noindent \textbf{Proof of Theorem \ref{Theorem 3.1} }: Consider the following probability, under $H_{1}^{*}$ (See, Equation \ref{equation:3.2}):
\begin{equation} \label{Critical eq}
    \begin{split}
        \mathds{P}_{H_{1}^{*}} \left[T_{n}>c\right] &= \mathds{P}_{H_{1}^{*}}\left[ \sum\limits_{i=0}^{d-1}\left|\hat{\beta}_{i,n}- \beta_{i,0}\right|>c\right]\\
 &\geq \mathds{P}_{H_{1}^{*}}\left[ \sum\limits_{i=0}^{d-1}\left(\hat{\beta}_{i,n}- \beta_{i,o}\right)>c\right]\\
 &= \mathds{P}_{H_{1}^{*}}\left[ \sum\limits_{i=0}^{d-1}\hat{\beta}_{i,n}>c+\sum\limits_{i=0}^{d-1}{\beta}_{i,0}\right]\\
 &= \mathds{P}_{H_{1}^{*}}\left[ \hat\beta_{0,n} +  \sum\limits_{i=1}^{d-1}\hat{\beta}_{i,n}>c+M\right]\\
 &=\mathds{P}_{H_{1}^{*}}\left[ \hat\beta_{0,n} + \sum\limits_{i=1}^{d-1}\left(\hat{\beta}_{i,n} -\mathbb{E}\left(\hat{\beta}_{i,n}\right)\right) >c+M - \sum\limits_{i=1}^{d-1}\mathbb{E}\left(\hat{\beta}_{i,n}\right) \right]
    \end{split}
\end{equation} 
where M = $\displaystyle{\sum\limits_{i=0}^{d-1}\beta_{i,0}}$ is a positive constant. Now, using the Lemma \ref{Lemma 2}, we have $$\eta_{1} < \mathbb{E}\left(\frac{\hat\beta_{i,n}}{n}\right) < \eta_{2} \text{ for all } n > n_{0},$$ which implies that for all $n > n_{0}$ we have 
\begin{equation} \label{equation lemma 2}
    \frac{c + M}{n} - \left(d-1\right)\eta_{2} \leq \frac{c + M}{n}- \frac{\sum\limits_{i=1}^{d-1} \mathbb{E} \left ( \hat {\beta}_{i,n} \right)}{n} \leq\frac{c + M}{n}-\left(d-1\right)\eta_{1}
\end{equation}

Thus, using the fact that for any random variable X and real numbers $a, b$ such that $a < b$, we have $\mathbb{P} \left(X > a\right) \geq \mathbb{P} \left(X > b\right)$. Therefore, using the Equation \ref{Critical eq} and Equation \ref{equation lemma 2}, we have the following: 
\begin{equation}\label{ equation 1 continued }
    \begin{split}
        \mathds{P}_{H_{1}^{*}} \left[T_{n}>c\right] 
        &\geq \mathds{P}_{H_{1}^{*}} \left[\frac{\hat\beta_{0,n}}{n} + \frac{\sum\limits_{i=1}^{d-1}\left(\hat{\beta}_{i,n} -\mathbb{E}\left(\hat{\beta}_{i,n}\right)\right)}{n} > \frac{c + M}{n}  - \frac{\sum\limits_{i=1}^{d-1}\mathbb{E}\left(\hat{\beta}_{i,n}\right)}{n}\right]\\ 
        &\geq\mathds{P}_{H_{1}^{*}} \left[\frac{\hat\beta_{0,n}}{n} + \frac{\sum\limits_{i=1}^{d-1}\left(\hat{\beta}_{i,n} -\mathbb{E}\left(\hat{\beta}_{i,n}\right)\right)}{n}-\frac{c + M}{n} > - \left(d-1\right)\eta_{1}\right]  
    \end{split}
\end{equation}

Now, we shall find the asymptotic distribution of  
$ \displaystyle{ \frac{\hat\beta_{0,n}}{n} + \sum\limits_{i=1}^{d-1}\frac{U_{i}}{n}-\frac{c + M}{n}}$, where for all $i = 1,\ldots d-1$, $\displaystyle{U_{i} = \hat{\beta}_{i,n} -\mathbb{E}\left(\hat{\beta}_{i,n}\right)}.$ We proceed in the following two steps:

\noindent\textbf{First}, using the Lemma \ref{Lemma 1} and Lemma \ref{Lemma 3} for $0< \xi < 1$ and for all $i = 1,\ldots d-1$, we have
\begin{equation} \label{ equation First}
    \frac{\hat\beta_{0,n}}{n} \xrightarrow{ a.s } \xi \Longrightarrow \frac{\hat\beta_{0,n}}{n} \xrightarrow{ D } \xi \text{ as } n \xrightarrow{} \infty
\end{equation}
\& \begin{equation} \label{equation a.s}
    \frac{U_{i}}{n} \xrightarrow{a.s} 0 \Longrightarrow \sum\limits_{i=1}^{d-1} \frac{U_{i}}{{n}} \xrightarrow{a.s} 0
\end{equation}

\noindent\textbf{Second}, note that since $c + M$ is a positive constant, $\frac{c + M}{n} \xrightarrow{} 0$ as $n \xrightarrow{} \infty$, therefore, from the Equation \ref{equation a.s}, we have
\begin{equation} \label{ equation 6}
    \sum\limits_{i=1}^{d-1}\frac{U_{i}}{n}-\frac{c + M}{n} \xrightarrow{ a.s} 0
\end{equation}

Therefore, using Slutsky's theorem with the Equation \ref{ equation First} and Equation \ref{ equation 6}, we have
\begin{equation} \label{equation 7}
   \frac{\hat\beta_{0,n}}{n} + \sum\limits_{i=1}^{d-1}\frac{U_{i}}{n}-\frac{c + M}{n} \xrightarrow{ D } U \text{ as } n \xrightarrow{} \infty
\end{equation}
where $U = \xi$ with probability 1. Since $ 0 < \xi < 1$, $U$ is a positive random variable. 

Therefore, from the Equation \ref{equation 7} and Equation \ref{ equation 1 continued }, we have 
$$ \begin{aligned}
    \mathds{P}_{H_{1}^{*}} \left[T_{n}>c\right] 
    &\geq\mathds{P}_{H_{1}^{*}} \left[\frac{\hat\beta_{0,n}}{n} + \frac{\sum\limits_{i=1}^{d-1}\left(\hat{\beta}_{i,n} -\mathbb{E}\left(\hat{\beta}_{i,n}\right)\right)}{n}-\frac{c + M}{n} > - \left(d-1\right)\eta_{1}\right]\\
    &\xrightarrow{} \mathds{P}_{H_{1}^{*}} \left[U > -\left(d-1\right)\eta_{1} \right]\\
    & = 1.
\end{aligned}$$
where the last line follows from the fact that $U$ is a positive random variable, $d \geq 1$ and $\eta_{1} > 0.$ Hence, consistency of the test is established.

\hfill$\Box$

\noindent \textbf{Proof of Theorem \ref{Theorem 3.2} }: Consider the following probability under $H_{1}^{*}$ (See, Equation \ref{equation:3.2} ):
\begin{equation} \label{ConsistencyP}
    \begin{split}
        \mathds{P}_{H_{1}^{*}} \left[T_{n}>c\right] &= \mathds{P}_{H_{1}^{*}}\left[ \sum\limits_{i=0}^{d-1}\left|\hat{\beta}_{i,n}- \beta_{i,0}\right|>c\right]\\
 &\geq \mathds{P}_{H_{1}^{*}}\left[ \sum\limits_{i=0}^{d-1}\left(\hat{\beta}_{i,n}- \beta_{i,o}\right)>c\right]\\
 &= \mathds{P}_{H_{1}^{*}}\left[ \sum\limits_{i=0}^{d-1}\hat{\beta}_{i,n}>c+\sum\limits_{i=0}^{d-1}{\beta}_{i,0}\right]\\
 &= \mathds{P}_{H_{1}^{*}}\left[ \hat\beta_{0,n} +  \sum\limits_{i=1}^{d-1}\hat{\beta}_{i,n}>c+M\right]\\
 &= \mathds{P}_{H_{1}^{*}} \left[\frac{\hat\beta_{0,n}}{n} + \frac{\sum\limits_{i=1}^{d-1}\hat{\beta}_{i,n}}{n} > \frac{c + M}{n}\right]
    \end{split}
\end{equation} 
Here, M = $\displaystyle{\sum\limits_{i=0}^{d-1}\beta_{i,0}}$ is a positive constant. Now, we shall find the asymptotic distribution of $\displaystyle{\frac{\hat\beta_{0,n}}{n} + \frac{\sum\limits_{i=1}^{d-1}\hat{\beta}_{i,n}}{n}}$ in the following \textbf{two steps}:

\textbf{First}, note that under the Assumption (A.2), from the Lemma \ref{Lemma 1} we have $\hat{\beta}_{0,n}$ asymptotically almost surely bounded $i.e$ there exist constants  $0<k<l<\infty$ such that:

$$ \mathds{P}\left[\frac{\hat\beta_{0,n}}{n}\geq k\right]\xrightarrow{} \text{ 1 as n } \xrightarrow{} \infty \text{ and } \mathds{P}\left[\frac{\hat\beta_{0,n}}{n}\leq l\right]\xrightarrow{}\text{ 1 as n } \xrightarrow{} \infty $$.

Therefore, for any $t\in \mathds{R}$, we have: 
$$\mathds{P}\left[\frac{\hat\beta_{0,n}}{n} \leq t\right]\xrightarrow{} \begin{cases}
    0,  & t < k \\
    p,  & k \leq t < l \\
    1,  & t \geq l
    \end{cases} \text{ , where p } \in ( 0, 1).$$

Hence, we have: 

\begin{equation} \label{E1}
    \mathds{P}\left[\frac{\hat\beta_{0,n}}{n} \leq t\right]\xrightarrow{} \mathds{P}\left[V \leq t\right] \Longrightarrow \frac{\hat\beta_{0,n}}{n} \xrightarrow{D} \text{ V as n }\xrightarrow{} \infty.
\end{equation} 

\textbf{Second}, note that under the Assumption (A.1) from the Lemma \ref{Lemma 2} we have: 
 
$$\mathds{E}\left(\hat\beta_{i,n}\right) = o(n) \text{ for all i = 1,}\ldots, d-1.$$

Since $\hat\beta_{i,n}$'s are non-negative random variables, using the Lemma \ref{Lemma 2}, we have:

$$\mathds{E}\left(\hat\beta_{i,n}\right) = o(n) \Longrightarrow{} \mathds{E}\left|\frac{\hat\beta_{i,n}}{n}\right|\xrightarrow{}0 \Longrightarrow \frac{\hat\beta_{i,n}}{n}\xrightarrow{L^{1}} \text{ 0 as n } \xrightarrow{}\infty.$$

As convergence in mean implies convergence in probability, we have $\displaystyle{\frac{\hat\beta_{i,n}}{n} \xrightarrow{P} 0.}$

Now, since $\displaystyle{\frac{\hat\beta_{i,n}}{n} \xrightarrow{P} \text{ 0 for all i = 1,}\ldots d-1}$ and $d$ is a fixed integer, we have: 

\begin{equation} \label{E2}
    \sum\limits_{i=1}^{d-1}\frac{\hat{\beta}_{i,n}}{n}\xrightarrow{P}\text{ 0 as n }\xrightarrow{} \infty.
\end{equation}

Next, using Slutsky's theorem with the Equation (\ref{E1}) and Equation (\ref{E2}), we have:

\begin{equation}\label{V}
    \frac{\hat\beta_{0,n}}{n} + \frac{\sum\limits_{i=1}^{d-1}\hat{\beta}_{i,n}}{n}\xrightarrow{D}  \text{V as n} \xrightarrow{} \infty.
\end{equation}
Now, using the Equation \ref{ConsistencyP} and Equation \ref{V}, we have:

$$
\begin{aligned}
    \mathds{P}_{H_{1}^{*}} \left[T_{n}>c\right]
    & \geq \mathds{P}_{H_{1}^{*}} \left[\frac{\hat\beta_{0,n}}{n} + \frac{\sum\limits_{i=1}^{d-1}\hat{\beta}_{i,n}}{n} > \frac{c + M}{n}\right]\\
    &\xrightarrow{} \mathds{P}_{H_{1}^{*}} \left[V > 0\right]\\
    & = 1,
    \end{aligned}
$$
where, the last line follows from the fact that $V$ is a positive random variable and since $\frac{c+M}{n}\xrightarrow{} 0 \text{ as n } \xrightarrow{}\infty$, using Slutsky's theorem we have:
$$\frac{\hat\beta_{0,n}}{n} + \frac{\sum\limits_{i=1}^{d-1}\hat{\beta}_{i,n}}{n} - \frac{c+M}{n}\xrightarrow{D}\text{V as n}\xrightarrow{} \infty.$$
Hence, consistency of the test based on $T_{n}$ is established.

\hfill$\Box$
 
 \noindent \textbf{Proof of Theorem \ref{ Theorem 3.3 } }: Consider the following probability, under $H_{1}^{**}$ (See, Equation \ref{equation 3.3}):
\begin{equation}\label{critical Two sample}
    \begin{split} 
        \mathds{P}_{H_{1}^{**}} \left[T_{n_{1}, n_{2}} > c \right] &= \mathds{P}_{H_{1}^{**}}\left[ \sum\limits_{i = 0}^{d -1}\left|\hat{\beta}_{i, n_{1}} - \hat{\beta}_{i, n_{2}}^{*}\right|>c\right]\\
        &\geq \mathds{P}_{H_{1}^{**}}\left[ \sum\limits_{i=0}^{d-1}\left(\hat{\beta}_{i, n_{1}} - \hat{\beta}_{i, n_{2}}^{*}\right)>c\right]\\
 &= \mathds{P}_{H_{1}^{**}}\left[ \hat{\beta}_{0, n_{1}}+\sum\limits_{i=1}^{d-1}\hat{\beta}_{i,n_{1}} - \hat{\beta}_{0, n_{2}}^{*} - \sum\limits_{i=1}^{d-1}\hat{\beta}_{i,n_{2}}^{*} > c\right]\\
 &= \mathds{P}_{H_{1}^{**}}\left[ \frac{\hat{\beta}_{0, n_{1}}+\sum\limits_{i=1}^{d-1}\hat{\beta}_{i,n_{1}}}{n_{1}+n_{2}} - \frac{\hat{\beta}_{0, n_{2}}^{*} + \sum\limits_{i=1}^{d-1}\hat{\beta}_{i,n_{2}}^{*}}{n_{1}+n_{2}} > \frac{c}{n_{1}+n_{2}}\right]\\
 &= \mathds{P}_{H_{1}^{**}}\left[ \left(\frac{n_{1}}{n_{1}+n_{2}}\right)W_{1,n_{1}} - \left(\frac{n_{2}}{n_{1}+n_{2}}\right)W_{2,n_{2}} > \frac{c - a_{n_{1}}-b_{n_{2}}}{n_{1}+n_{2}}\right],\\
 \end{split}
\end{equation}
 where $\displaystyle{W_{1,n_{1}} =  \frac{\hat{\beta}_{0, n_{1}}+\sum\limits_{i=1}^{d-1}\left(\hat{\beta}_{i,n_{1}} -\mathbb{E}\left(\hat{\beta}_{i,n_{1}}\right)\right)}{n_{1}}}$, 
 $\displaystyle{W_{2,n_{2}} =  \frac{\hat{\beta}_{0, n_{2}}+\sum\limits_{i=1}^{d-1}\left(\hat{\beta}_{i,n_{2}} -\mathbb{E}\left(\hat{\beta}_{i,n_{2}}\right)\right)}{n_{2}}}$,
 
 \noindent $\displaystyle{ a_{n_{1}} = \sum\limits_{i=1}^{d-1} \mathbb{E}\left(\hat{\beta}_{i,n_{1}}\right)}$ and $\displaystyle{ b_{n_{2}} = \sum\limits_{i=1}^{d-1} \mathbb{E}\left(\hat{\beta}_{i,n_{2}}\right)}$.
 
 Now, using the Lemma \ref{Lemma 2} for the random samples  $\mathbb{X}_{n_{1}}$ and $\mathbb{Y}_{n_{2}}$, ( See, Section \ref{ Theorem 3.3 } ) we have the following:
 \begin{enumerate}[(i)]
     \item $\displaystyle{n_{1}\theta_{1} < \mathbb{E}\left(\hat\beta_{i,n_{1}}\right) < n_{1}\theta_{2} \text{ for all } n_{1} > N_{0}}$
     \item $\displaystyle{n_{2}\zeta_{1} < \mathbb{E}\left(\hat\beta_{i,n_{2}}\right) < n_{2}\zeta_{2} \text{ for all } n_{2} > M_{0} }$
 \end{enumerate}

\noindent Further, using (i) and (ii), we have the following two inequalities:
\begin{equation} \label{Equation 9}
   c - n_{1} \theta_{2}\left(d-1\right)\leq c -  \sum\limits_{i=1}^{d-1} \mathbb{E}\left(\hat{\beta}_{i, n_{1}}\right) \leq c - n_{1} \theta_{1}\left(d-1\right)
\end{equation}
\& \begin{equation} \label{Equation 10}
   -n_{2} \zeta_{2}\left(d-1\right)\leq  -  \sum\limits_{i=1}^{d-1} \mathbb{E}\left(\hat{\beta}_{i, n_{2}}\right) \leq  - n_{2} \zeta_{1}\left(d-1\right)
\end{equation}

Now, using the Equation \ref{Equation 9} and Equation \ref{Equation 10}, we have:
\begin{equation} \label{equation 11}
    c - n_{1} \theta_{2} \left(d-1\right)-n_{2} \zeta_{2}\left(d-1\right)\leq c - a_{n_{1}} - b_{n_{2}}\leq c - n_{1} \theta_{1}\left(d-1\right)-n_{2} \zeta_{1}\left(d-1\right)
\end{equation}

Thus, using the fact that for any random variable X and real numbers $a, b$ such that $a < b$, we have $\mathbb{P} \left(X > a\right) \geq \mathbb{P} \left(X > b\right)$. Therefore, using the Equation \ref{critical Two sample} and Equation \ref{equation 11}, we have the following:
\begin{equation} \label{Equation 12}
    \begin{split}
        \mathds{P}_{H_{1}^{**}} \left[T_{n_{1}, n_{2}} > c \right]
        &\geq \mathds{P}_{H_{1}^{**}}\left[\left(\frac{n_{1}}{n_{1}+n_{2}}\right)W_{1,n_{1}} - \left(\frac{n_{2}}{n_{1}+n_{2}}\right)W_{2,n_{2}} > \frac{c - \left(d-1\right)\left(n_{1} \theta_{1} + n_{2} \zeta_{1}\right)}{n_{1}+n_{2}}\right]
    \end{split}
\end{equation}
Now, we shall derive the asymptotic distribution of $W_{1,n_{1}}$ and $W_{2,n_{2}}$ in the following two steps:

\noindent\textbf{First}, using the Lemma \ref{Lemma 1} and Lemma \ref{Lemma 3} for the random sample $\mathbb{X}_{n_{1}}$, we have the following:
\begin{equation} \label{ Equation 13}
    \frac{\hat\beta_{0,n_{1}}}{n_{1}} \xrightarrow{ a.s } \theta \Longrightarrow \frac{\hat\beta_{0,n_{1}}}{n_{1}} \xrightarrow{ D } \theta \text{ as } n_{1} \xrightarrow{} \infty
\end{equation}
\& \begin{equation} \label{Equation 14}
     \frac{\sum\limits_{i=1}^{d-1}\left(\hat{\beta}_{i, n_{1}} - \mathbb{E}\left(\hat{\beta}_{i, n_{1}}\right)\right)}{{n_{1}}} \xrightarrow{a.s} 0
\end{equation} 
Therefore, using Slutsky's theorem with the Equation \ref{ Equation 13} and Equation \ref{Equation 14} we have 
\begin{equation} \label{Equation 15}
      \frac{\hat{\beta}_{i,n_{1}} +\sum\limits_{i=1}^{d-1}\left(\hat{\beta}_{i, n_{1}} - \mathbb{E}\left(\hat{\beta}_{i, n_{1}}\right)\right)}{{n_{1}}} \xrightarrow{D} \theta \Longrightarrow W_{1,n_{1}} \xrightarrow{D} W
\end{equation}
where $W = \theta$ with probability 1. Since $0 < \theta < 1$ from the Lemma \ref{Lemma 1}, $W$ is a positive random variable. 

\noindent \textbf{Second}, using the Lemma \ref{Lemma 1} and Lemma \ref{Lemma 3} for the random sample $\mathbb{X}_{n_{2}}$ and following the same procedure as in the first step, we have
\begin{equation} \label{ Equation 16}
    \frac{\hat{\beta}_{i,n_{1}} +\sum\limits_{i=1}^{d-1}\left(\hat{\beta}_{i, n_{2}} - \mathbb{E}\left(\hat{\beta}_{i, n_{2}}\right)\right)}{{n_{2}}} \xrightarrow{D} \zeta \Longrightarrow W_{2,n_{2}} \xrightarrow{D} W^{'}
\end{equation}
where $W^{'} = \zeta$ with probability 1. Since $0 < \zeta < 1$ from the Lemma \ref{Lemma 1}, $W^{'}$ is a positive random variable. 

Now, using the Equation \ref{Equation 12} with the Equation \ref{Equation 15} and Equation \ref{ Equation 16} and using the fact that $\displaystyle{\frac{n_{1}}{n_{1} + n_{2}} \xrightarrow{} \lambda \in \left(0, \infty\right) \text{ as min}\left(n_{1}, n_{2}\right) \xrightarrow{} \infty}$, we have
$$\begin{aligned}
   \mathds{P}_{H_{1}^{**}} \left[T_{n_{1}, n_{2}} > c \right]
        &\xrightarrow{}\mathds{P}_{H_{1}^{**}}\left[\lambda W - \left(1 - \lambda\right)W^{'} > - \left(d-1\right)\left(\lambda \theta_{1} + \left(1 - \lambda\right)\zeta_{1}\right)\right]\\
        &= \mathds{P}_{H_{1}^{**}} \left[\lambda W > \left(1 - \lambda\right)\zeta - \left(d-1\right)\left(\lambda \theta_{1} + \left(1 - \lambda\right)\zeta_{1}\right) \right]\\
        &=1 \text{ as min}\left(n_{1}, n_{2}\right) \xrightarrow{} \infty
\end{aligned}$$
where the last line follows from the fact that $W$ and $W^{'}$ are independent positive random variables and assuming a mild condition that $\displaystyle{\lambda\theta_{1} > \left(1 - \lambda\right)\zeta}$. 

\noindent This establishes consistency of the two sample test under the critical regime.

\hfill$\Box$

\noindent \textbf{Proof of Theorem \ref{Theorem 3.4}:} Consider the following probability under $H_{1}^{**}$ (See, Equation \ref{equation 3.3}):
\begin{equation}\label{Probability RHS Two sample}
    \begin{split}
        \mathds{P}_{H_{1}^{**}} \left[T_{n_{1}, n_{2}} > c \right] &= \mathds{P}_{H_{1}^{**}}\left[ \sum\limits_{i = 0}^{d -1}\left|\hat{\beta}_{i, n_{1}} - \hat{\beta}_{i, n_{2}}^{*}\right|>c\right]\\
        &\geq \mathds{P}_{H_{1}^{**}}\left[ \sum\limits_{i=0}^{d-1}\left(\hat{\beta}_{i, n_{1}} - \hat{\beta}_{i, n_{2}}^{*}\right)>c\right]\\
 &= \mathds{P}_{H_{1}^{**}}\left[ \hat{\beta}_{0, n_{1}}+\sum\limits_{i=1}^{d-1}\hat{\beta}_{i,n_{1}} - \hat{\beta}_{0, n_{2}}^{*} - \sum\limits_{i=1}^{d-1}\hat{\beta}_{i,n_{2}}^{*} > c\right]\\
 &= \mathds{P}_{H_{1}^{**}}\left[ \frac{\hat{\beta}_{0, n_{1}}+\sum\limits_{i=1}^{d-1}\hat{\beta}_{i,n_{1}}}{n_{1}+n_{2}} - \frac{\hat{\beta}_{0, n_{2}}^{*} + \sum\limits_{i=1}^{d-1}\hat{\beta}_{i,n_{2}}^{*}}{n_{1}+n_{2}} > \frac{c}{n_{1}+n_{2}}\right]\\
 &= \mathds{P}_{H_{1}^{**}}\left[ \left(\frac{n_{1}}{n_{1}+n_{2}}\right)Z_{1} - \left(\frac{n_{2}}{n_{1}+n_{2}}\right)Z_{2} > \frac{c}{n_{1}+n_{2}}\right],\\
 \end{split}
\end{equation}
 where $\displaystyle{Z_{1}= \frac{\hat{\beta}_{0, n_{1}}+\sum\limits_{i=1}^{d-1}\hat{\beta}_{i,n_{1}}}{n_{1}}}$ and 
 $\displaystyle{Z_{2}=\frac{\hat{\beta}_{0, n_{2}}^{*} + \sum\limits_{i=1}^{d-1}\hat{\beta}_{i,n_{2}}^{*}}{n_{2}}}$.

 Now, under the Assumptions (A.1) and (A.2), we will find the asymptotic distribution of $Z_{1}$ and $Z_{2}$ using the Lemma \ref{Lemma 1} and Lemma \ref{Lemma 2}. Note that since $\mathbb{X}$ and $\mathbb{Y}$ are two independent samples, therefore using the Lemma \ref{Lemma 1} and Lemma \ref{Lemma 2} for $\mathbb{X}$ and $\mathbb{Y}$, in the same line of approach as in Theorem \ref{Theorem 3.1} (See Equation \ref{V}), we have the following:
 \begin{enumerate}[(i)]
     \item $\displaystyle{\frac{\hat\beta_{0,n_{1}}}{n_{1}} + \frac{\sum\limits_{i=1}^{d-1}\hat{\beta}_{i,n_{1}}}{n_{1}}\xrightarrow{D} V_{1}\text{ as } n_{1}\xrightarrow{} \infty} \text{ i.e } Z_{1} \xrightarrow{D} V_{1} \text{ as } n_{1}\xrightarrow{} \infty.$
    
    \item $\displaystyle{\frac{\hat\beta_{0,n_{2}}^{*}}{n_{2}} + \frac{\sum\limits_{i=1}^{d-1}\hat{\beta}_{i,n_{2}}^{*}}{n_{2}}\xrightarrow{D} V_{2}\text{ as } n_{2} \xrightarrow{} \infty} \text{ i.e } Z_{2} \xrightarrow{D} V_{2} \text{ as } n_{2}\xrightarrow{} \infty.$ 
 \end{enumerate}

\noindent Here $V_{j}$'s, $j = 1,2$, are independent positive random variables defined as follows:
$$
V_{j} = \begin{cases} 
        k_{j}, \text{ with probability } p_{j}\\
        l_{j}, \text{ with probability } 1-p_{j}
 \end{cases}, \text{ where } p_{j} \in (0, 1) \text{ \& } 0<k_{j}<l_{j}<\infty.
$$
Also, note that since $\displaystyle{\frac{n_{1}}{n_{1}+n_{2}} \xrightarrow{} \lambda \in \left(0,1\right) \text{ as min}\left(n_{1}, n_{2}\right) \xrightarrow{} \infty}$, using (i) above, we have:
\begin{equation}\label{Eq7.10}
    \left(\frac{n_{1}}{n_{1}+n_{2}}\right)Z_{1} \xrightarrow{D} \lambda V_{1}, \text{ as min}\left(n_{1}, n_{2}\right) \xrightarrow{} \infty.
    \end{equation}
Similarly, using (ii) we have:
\begin{equation}\label{Eq7.11}
    \left(\frac{n_{2}}{n_{1}+n_{2}}\right)Z_{2} \xrightarrow{D} \left(1 - \lambda\right)V_{2}, \text{ as min}\left(n_{1}, n_{2}\right) \xrightarrow{} \infty.
    \end{equation}

Now, observe that $\displaystyle{\frac{c}{n_{1}+n_{2}}\xrightarrow{} 0 \text{ as min}\left(n_{1}, n_{2}\right) \xrightarrow{} \infty }$ and since $V_{1}$ and $V_{2}$ are independent, using Equation \ref{Eq7.10} and Equation \ref{Eq7.11}, as min($n_{1},n_{2}$) $\xrightarrow{}\infty$ we have:

\begin{equation}\label{Eq7.12}
  \left(\frac{n_{1}}{n_{1}+n_{2}}\right)Z_{1} - \left(\frac{n_{2}}{n_{1}+n_{2}}\right)Z_{2} - \frac{c}{n_{1}+n_{2}} \xrightarrow{D} \lambda V_{1} - \left(1 - \lambda\right)V_{2} 
\end{equation}

Now, using the Equation \ref{Probability RHS Two sample} and Equation \ref{Eq7.12}, we have:
$$
\begin{aligned}
   \mathds{P}_{H_{1}^{**}} \left[T_{n_{1}, n_{2}} > c \right]
   &\geq \mathds{P}_{H_{1}^{**}}\left[ \left(\frac{n_{1}}{n_{1}+n_{2}}\right)Z_{1} - \left(\frac{n_{2}}{n_{1}+n_{2}}\right)Z_{2} > \frac{c}{n_{1}+n_{2}}\right]\\
   &\xrightarrow{}\mathds{P}_{H_{1}^{**}}\left[\lambda V_{1} - \left(1-\lambda\right)V_{2} > 0 \right ] \text{ as min}\left(n_{1}, n_{2}\right)\xrightarrow{} \infty\\.
 &=\mathds{P}_{H_{1}^{**}}\left[V_{1} > \frac{\left(1-\lambda\right)V_{2}}{\lambda}\right]\\
 &= \mathds{P}_{H_{1}^{**}}\left[V_{1} > \frac{\left(1-\lambda\right)k_{2} }{\lambda} \right]p_{2} + \mathds{P}_{H_{1}^{**}}\left[V_{1} > \frac{\left(1-\lambda\right) l_{2}}{\lambda}\right]\left(1 - p_{2}\right)\\
 &= 1,
\end{aligned}
$$
where the last line follows from the fact that $V1$ and $V2$ are independent and assuming a mild condition that $\displaystyle{\left(1 - \lambda \right)l_{2} < \lambda k_{1}}$ which implies that: $$\text{ max}\left(\frac{\left(1-\lambda\right)k_{2} }{\lambda},\frac{\left(1-\lambda\right) l_{2}}{\lambda}\right) < k_{1}.$$
Thus, using the fact that $\displaystyle{\mathds{P}\left[V_{1} > t\right] = 1}$ for all $t < k_{1}$, consistency of the test based on $T_{n_{1},n_{2}}$ is established.

Hence, consistency of the test based on $T_{n_{1},n_{2}}$ is established.
 
 \hfill$\Box$

\section*{Acknowledgments}
Both authors are thankful to Shamriddha De (presently a PhD (Statistics) student at the University of Iowa, USA; a former M.Sc. (Statistics) student at the IIT Kanpur) for his initial involvement when the authors started learning topological data analysis. The second author is partially supported by a CRG grant (CRG/2022/001489), a research grant from the SERB, Government of India.

\bibliographystyle{dinat}
\bibliography{main} 
\end{document}